\documentclass{aa}
\usepackage[varg]{txfonts}
\usepackage[font=footnotesize,labelfont={bf,sc,small},figurename=Fig.]{caption}
\usepackage[hidelinks]{hyperref}
\graphicspath{{../figures/}}
\usepackage{subfig}
\usepackage{rotate}
\usepackage{appendix}
\usepackage{xcolor}
\usepackage{varioref}
\labelformat{figure}{Fig.~#1}
\labelformat{section}{Sect.~#1}
\labelformat{subsection}{Sect.~#1}
\labelformat{subsubsection}{Sect.~#1}
\labelformat{appendix}{Appendix~#1}
\labelformat{table}{Table~#1}
\usepackage{natbib}
\usepackage{color}
\bibpunct{(}{)}{;}{a}{}{,} % to follow the A&A style
\hypersetup{draft}

\usepackage{multirow}

\begin{document}

\title{The X-ray activity of the young solar-like star Kepler-63 and the structure of its corona}
\author{M. Coffaro\inst{1} 
\and B. Stelzer\inst{\ref{inst1},\ref{inst2}}
\and S. Orlando\inst{2}
}

\offprints{M. Coffaro}

\institute{Institut f\"ur Astronomie und Astrophysik T\"ubingen, Kepler Center for Astro and Particle Physics, Eberhard Karls Universit\"at,  Sand 1, 72076 T\"ubingen,Germany \\
           \email{coffaro@astro.uni-tuebingen.de}\label{inst1}
\and INAF - Osservatorio Astronomico di Palermo, Piazza del Parlamento 1, I-90134 Palermo, Italy\label{inst2}
}

\date{Received 24 September 2021 / Accepted 11 February 2022}

\abstract{
The X-ray satellite \textit{XMM-Newton} has so far revealed coronal cycles in seven solar-like stars. In this sample, the youngest stars $\epsilon$~Eridani ($\sim 400$~Myr) and $\iota$~Horologii ($\sim 600$~Myr) display the shortest X-ray cycles and the smallest amplitudes, defined as the variation of the X-ray luminosity between the maximum and minimum of the cycle. The X-ray cycle of $\epsilon$~Eridani was characterized by applying a novel technique that allowed to model the corona of a solar-like star in terms of  magnetic structures observed on the Sun (active regions, cores of active regions and flares) at varying filling factors. The high surface coverage of $\epsilon$~Eridani with magnetic structures ($65-95\%$) that emerged from that study was held responsible for the low cycle amplitude in the X-ray band. It was also hypothesized that the basal surface coverage with magnetic structures may be higher on the corona of solar-like stars while they are young. To investigate this hypothesis, we started in 2019 the X-ray monitoring campaign of Kepler-63. The star had been observed once before, in 2014, with \textit{XMM-Newton}. With an age of $210 \pm 45$~Myr and a photospheric cycle of $1.27$~yr, Kepler-63 is so far the youngest star observed in X-rays with the aim of revealing a coronal cycle. Our campaign comprised four X-ray observations of Kepler-63, spanning 10 months (i.e. 3/5 of its photospheric cycle). In the long-term X-ray lightcurve we did not reveal a periodic variation of the X-ray 
luminosity,  but a factor two change would be compatible with the
considerable uncertainties in the low signal-to-noise data of this relatively
distant star.
%, compatible with the photospheric cycle, due to the low SN ratio of the X-ray 
%spectra. 
As for the case of $\epsilon$~Eridani, 
we described the coronal emission measure distribution (EMD) of Kepler-63 
with magnetic structures observed on the Sun. 
The best match with the observations is found for an EMD composed
of cores and flares of GOES Class C and M following the canonical flare frequency distribution. More energetic flares can be occasionally present but they do not contribute significantly to the quasi-stationary high-energy component of the emission measure probed with our modelling. This model yields a coronal filling factor
of $100\%$.
This complete coverage of the corona with X-ray emitting magnetic structures is
consistent with the absence of an X-ray cycle, confirming the analogous 
results derived earlier for $\epsilon$~Eridani. 
Finally, 
combining our results with the literature on stellar X-ray cycles we establish an empirical
relation between the cycle amplitude $L_{\rm X, max} / L_{\rm X, min}$ and the X-ray surface flux, $F_{\rm X,surf}$. From the absence of a coronal cycle in Kepler-63 we infer 
that stars with higher X-ray flux than Kepler-63 must host an EMD that comprises a significant fraction of higher-energetic flares than those 
necessary to model the corona of Kepler-63, i.e. flares of Class\,X or higher. 
Our study 
opens new ground for studies of the solar-stellar analogy and the joint exploration of resolved 
and unresolved variability in stellar X-ray lightcurves. }

\maketitle
\interfootnotelinepenalty=10000

\section{Introduction}
\label{sec:intro}

The magnetic field in solar-like stars is maintained by the so-called $\alpha-\Omega$ dynamo, that is responsible of changing the configuration of the magnetic field lines periodically. The time scale on which these changes take place defines the length of the activity cycle \citep{2007ApJ...657..486B}. A consequence of the magnetic field re-configuration is the formation of magnetic structures that rise on the stellar surface, evolve and decay periodically in the course of the magnetic cycle. The surface coverage with these structures, thus, varies continuously between a minimum and a maximum. Therefore, by tracing the evolution of the magnetic structures we can characterize the activity cycles of solar-like stars. These structures are manifest in all layers of the stellar atmosphere, from the photosphere to the corona. Due to the temperature gradient that characterizes the solar and stellar atmospheres,  triggering different emission mechanisms for each layer of the atmosphere, activity cycles are expected to be manifest at different wavelengths, from the optical to the X-rays.

Photospheric and chromospheric cycles are well studied. In particular, $\sim 60\%$ of the main-sequence stars show chromospheric cycles with periods lasting up to 20~yr \citep{1978ApJ...226..379W}. These cycles are characterized with Ca\,II H\&K emission lines emitted by plages in the chromosphere.  The photospheric cycles can instead be characterized by tracing the evolution of the starspots. One way of gathering information on the starspot evolution is the so-called spot transit method \citep{2003ApJ...585L.147S}, that infers the presence of starspots through the brightness changes during a planet-transit. 

Our knowledge about cycles in the outermost atmospheric layer, the corona,  is still scarce. Since the same dynamo process is responsible for the magnetic cycle in all parts of the stellar atmosphere, the length of X-ray cycles is expected to coincide with that of the chromospheric cycles. Since the latter have been seen that last up to 20~yr, long X-ray monitoring campaigns would be required to detect their coronal counterparts. Moreover, the search for X-ray cycles relies on sparsely sampled lightcurves. This is clearly a disadvantage for reliable coronal cycle detections. Therefore, studies of long-term X-ray variability have focused on stars with known chromospheric cycle.  

Up to date we are aware of the existence of only seven stars that show X-ray activity cycles, all detected by the X-ray satellite \textit{XMM-Newton}. For most of them, the common denominator is their age and the period of their coronal cycles. These stars are $\alpha$ Centauri A and B \citep{2012A&A...543A..84R}, $61$ Cygnus A and B \citep{2006A&A...460..261H, 2012A&A...543A..84R}, HD 81809 \citep{2008A&A...490.1121F, 2017A&A...605A..19O}, with ages between  $3$~Gyr and  $6$~Gyr and with X-ray activity cycles that last from $7$ up to $20$~yr. The remaining two stars with observed X-ray cycle, $\iota$~Horologii \citep{2013A&A...553L...6S, 2019A&A...631A..45S} and $\epsilon$~Eridani \citep{2020A&A...636A..49C}, are young stars, with age of 600 and 400~Myr and with short coronal cycles that have periods of $1.6$ and $2.9$~yr, respectively. 
Moreover, these two young stars show another intriguing characteristics: the amplitude of their X-ray cycles, i.e. the variation of the X-ray luminosity between the minimum and maximum of their coronal emission, is remarkably small. 

The X-ray emission of the Sun throughout its cycle is related to magnetic structures whose coronal coverage fraction changes along with the solar cycle. Exploiting the solar-stellar analogy, the hypothesis is that, for solar-like stars, we can link their X-ray variability to the same phenomenon. In this context, \citet{2020A&A...636A..49C} applied a novel technique which enables to reproduce the X-ray emission of a star in terms of time-variations in the coverage of its corona with the same kind of magnetic structures that are observed on the Sun. This technique was developed in the framework of the study "The Sun as an X-ray star" (\citealt{2000ApJ...528..524O}, \citealt{2000ApJ...528..537P}, \citealt{2001ApJ...557..906R}, \citealt{2001ApJ...560..499O}; see \ref{sec:sim_summary}) in which  observations of the solar corona by the X-ray satellite \textit{Yohkoh} were converted into a format virtually identical to that of the observations obtained with non-solar X-ray satellites, such as \textit{XMM-Newton}. The final data products of this method are synthetic X-ray spectra that are distinguished by a particular fractional coverage of magnetic structures on the corona, and can be directly compared to the observed spectra of a coronally active star: from such a comparison a percentage surface coverage with solar magnetic structures can be inferred for each stellar X-ray observation. 

This method was first applied to HD81809 \citep{2017A&A...605A..19O}. A more detailed study was carried out by \cite{2020A&A...636A..49C} for $\epsilon$~Eridani.
In that work we found a massive presence of magnetic structures on the corona throughout the X-ray activity cycle of $\epsilon$~Eridani. We argued that the high percentage (from $\sim 70$ to $\sim 90\%$) of the magnetic structures on the corona do not allow the X-ray luminosity to significantly vary during the coronal cycle, explaining the small amplitude of $\epsilon$~Eridani's X-ray luminosity. 

To better understand the origin of small cycle amplitudes and to investigate if solar-like stars younger than $\epsilon$~Eridani show X-ray activity cycles, we started the study of Kepler-63, a young G2V star at a distance of 200 pc. With its age of $210 \pm 45$~Myr \citep{2013ApJ...775...54S}, it is younger than the other stars that show short coronal activity cycles. It is also a fast rotator, with a rotational period of $\sim 5.4$~days and it hosts a Jupiter-like planet with a polar orbit of $9.43$~days \citep{2013ApJ...775...54S}.

Kepler-63 was extensively observed with the NASA satellite \textit{Kepler} and a photospheric cycle was detected by \citet{2016ApJ...831...57E} and \citet{2020A&A...635A..78N} from the analysis of the \textit{Kepler} lightcurves, by applying the spot transit method from which they found a significant periodicity at $1.27$~yr.
Its young age and the short activity cycle make Kepler-63 an ideal target for an X-ray monitoring campaign in search of its coronal cycle. 

In \ref{sec:obs_log} we present our X-ray monitoring campaign and the analysis of the X-ray observations. In \ref{sec:models}, we apply the method developed by \citet{ 2017A&A...605A..19O} and refined by \citet{2020A&A...636A..49C} with the aim to model the corona of Kepler-63 in terms of the same type of magnetic structures observed on our Sun. In \ref{sec:disc}, we discuss our results and we give our conclusions.

\section{Observations and data analysis}
\label{sec:obs_log}
The X-ray monitoring campaign of Kepler-63 with the satellite \textit{XMM-Newton} started in 2019 and it continued until 2020 (PI: M.Coffaro; Proposal ID: 084123). During this monitoring, four X-ray observations were acquired. The time span between each X-ray observation was of few months and the full X-ray campaign covered 3/5 of the photospheric cycle. 

Prior to this X-ray campaign, Kepler-63 had already been observed with \textit{XMM-Newton} once in 2014 \citep{2018MNRAS.477..808L}. This earlier observation was used to define the observational strategy of the  X-ray monitoring campaign. In \ref{tab:obs_log}, the \textit{XMM-Newton} observing log of Kepler-63 is presented, including all available observations. 

Kepler-63 was observed employing all instruments on board of \textit{XMM-Newton}. In this paper, only the analysis of the EPIC/pn data is presented. 

\begin{table}
\centering
\caption{Observing log of the X-ray monitoring campaign of Kepler-63.}
\begin{minipage}{0.5\textwidth}
\resizebox{\textwidth}{!}{
\begin{tabular}{cccc}
\hline
Obs. date & Obs ID & Exposure time\footnote{Time of acquisition of each observation of the EPIC/pn detector, prior to the GTI cuts.} & Count rate\footnote{EPIC/pn count rate obtained from the source detection procedure in the energy band $0.2-2.0$~keV.}\\
     &       &  [ksec]        &   [cnt/sec] \\
\hline
2014-09-28 & 0743460301\footnote{Archival observation} & $27$ & $0.0197 \pm 0.0011$\\
2019-05-05 & 0841230201 & $15.9$ & $0.0169 \pm 0.0020$ \\
2019-08-21 & 0841230301 & $15.7$ & $0.0170 \pm 0.0014$\\
2019-10-31 & 0841230401 & $9.5$ & $0.0185 \pm 0.0022$\\
2020-03-06 & 0841230501 & $5.7$ & $0.0197 \pm 0.0047$\\ 
\hline
\end{tabular}
}
\end{minipage}
\label{tab:obs_log}
\end{table}

For some observations the lightcurves of the EPIC/pn background showed irregularities and high count rates during the acquisition. Our visual inspection showed that the standard values for identifying the Good Time Intervals (GTIs), i.e. \texttt{RATE}~$< 0.4$~cnt/s as suggested by the threads of Science Analysis Software (\texttt{SAS}; version 1.3;  \citealt{SAS}) for the extraction of the EPIC/pn scientific data, did not remove all times of high background. Instead, we defined the GTIs \textit{ad  hoc}, identifying visually those intervals of time were the full-detector count rate was constant and low. In \ref{fig:bkg_lc}, the lightcurves of the full-detector background of the EPIC/pn are shown for the five observations. For the observations of September 2014 and August 2019, the GTIs were chosen over the whole lightcurve of the detector background, corresponding to \texttt{RATE}~$< 0.8$~cnt/s and \texttt{RATE}~$<1.45$~cnt/s respectively (red solid lines in \ref{fig:bkg_lc}). For the remaining observations, the GTIs were chosen as specific time intervals (highlighted by the red ranges in the plots). Because of this choice, these latter observations effectively result shorter in time with respect to their original exposure time.

After applying the source detection procedure of \texttt{SAS}, Kepler-63 results detected in all 5 EPIC/pn observations. 
 The positions of Kepler-63 on the detector identified by the source detection procedure were used to localize the extraction regions for the lightcurves and spectra: a circular region centered at the coordinates given by the source detection and with a radius of $20$~arcsec was chosen.
The background regions were chosen in a source-free part of the CCD where Kepler-63 was detected (and nearby the source). In the following subsections, the details on the lightcurve and spectral analysis are given.
\subsection{EPIC/pn lightcurves}
\label{sec:lc}
The EPIC/pn lightcurves of Kepler-63 were extracted in the soft energy band of \textit{XMM-Newton} ($0.2 - 2.0$~keV), employing the tools of \texttt{SAS}, \texttt{evselect} and \texttt{epiclccorr}. All lightcurves were binned with a time bin size of $1000$ seconds, with the exception of the observation of March 2020. In this case, the GTIs have a duration that is less than $1000$ seconds and, thus, a smaller bin size is more suitable. We chose $600$ seconds.

\begin{figure*}
\subfloat{\includegraphics[width=0.5\textwidth]{ 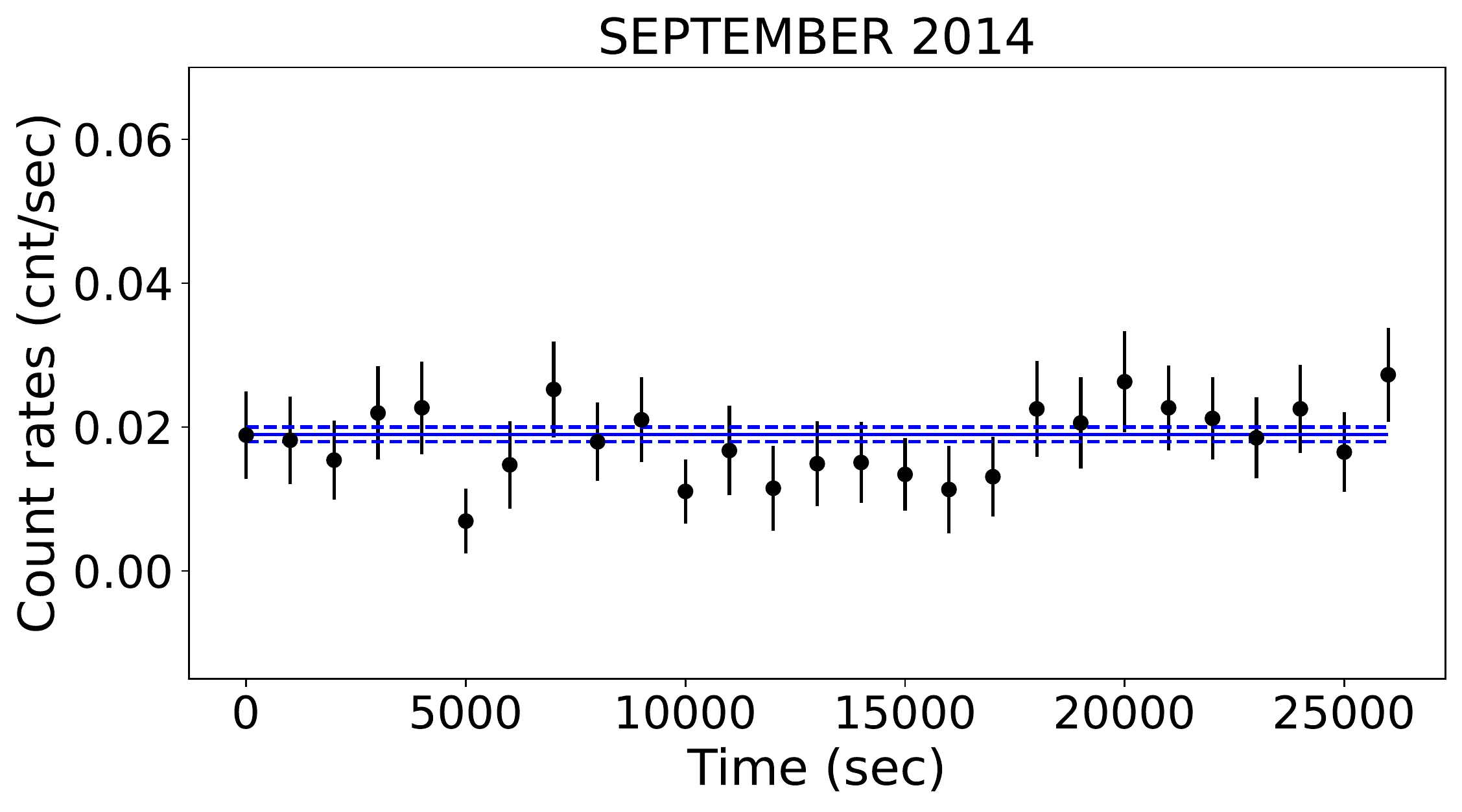}}
\subfloat{\includegraphics[width=0.5\textwidth]{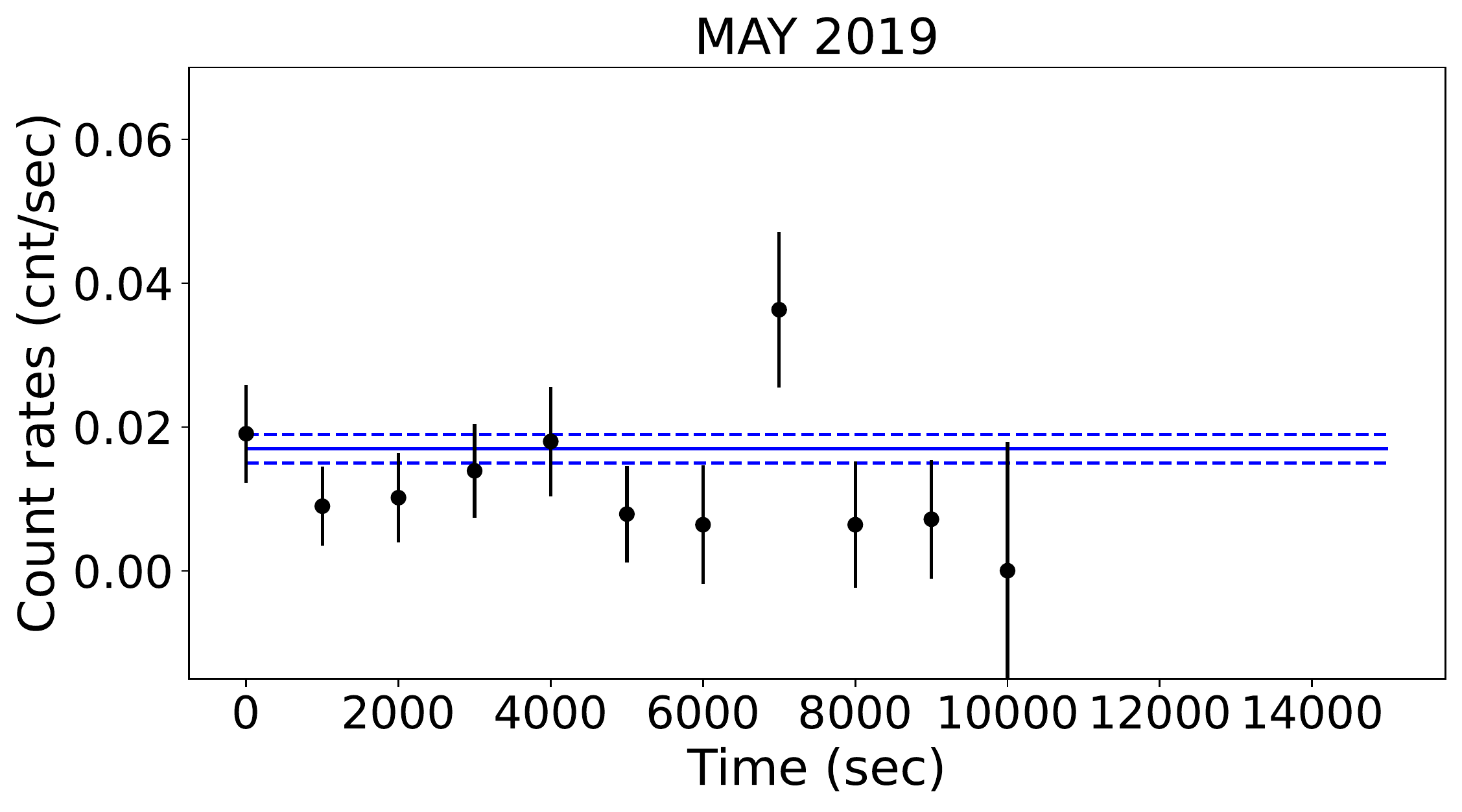}}\\
\subfloat{\includegraphics[width=0.5\textwidth]{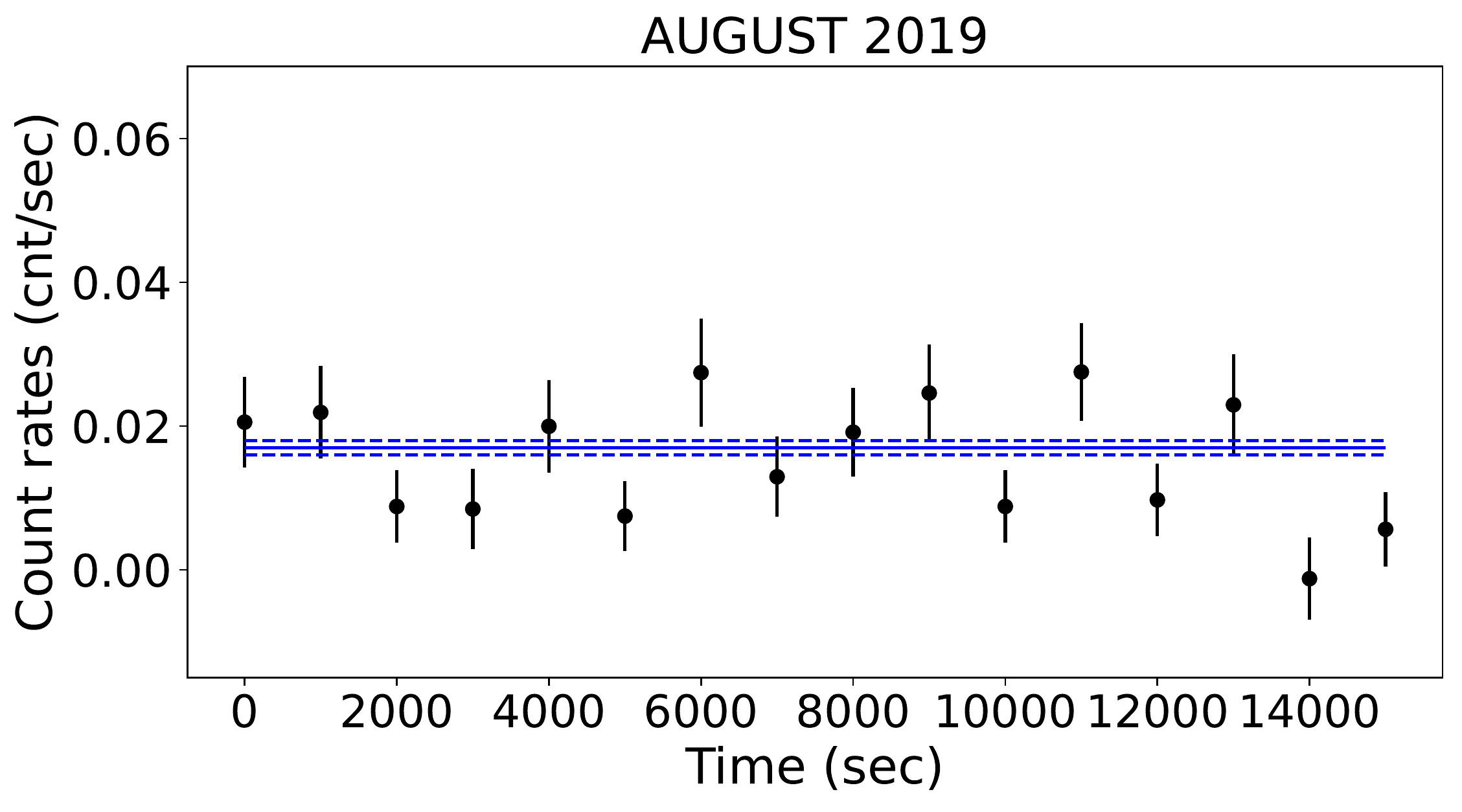}}
\subfloat{\includegraphics[width=0.5\textwidth]{ 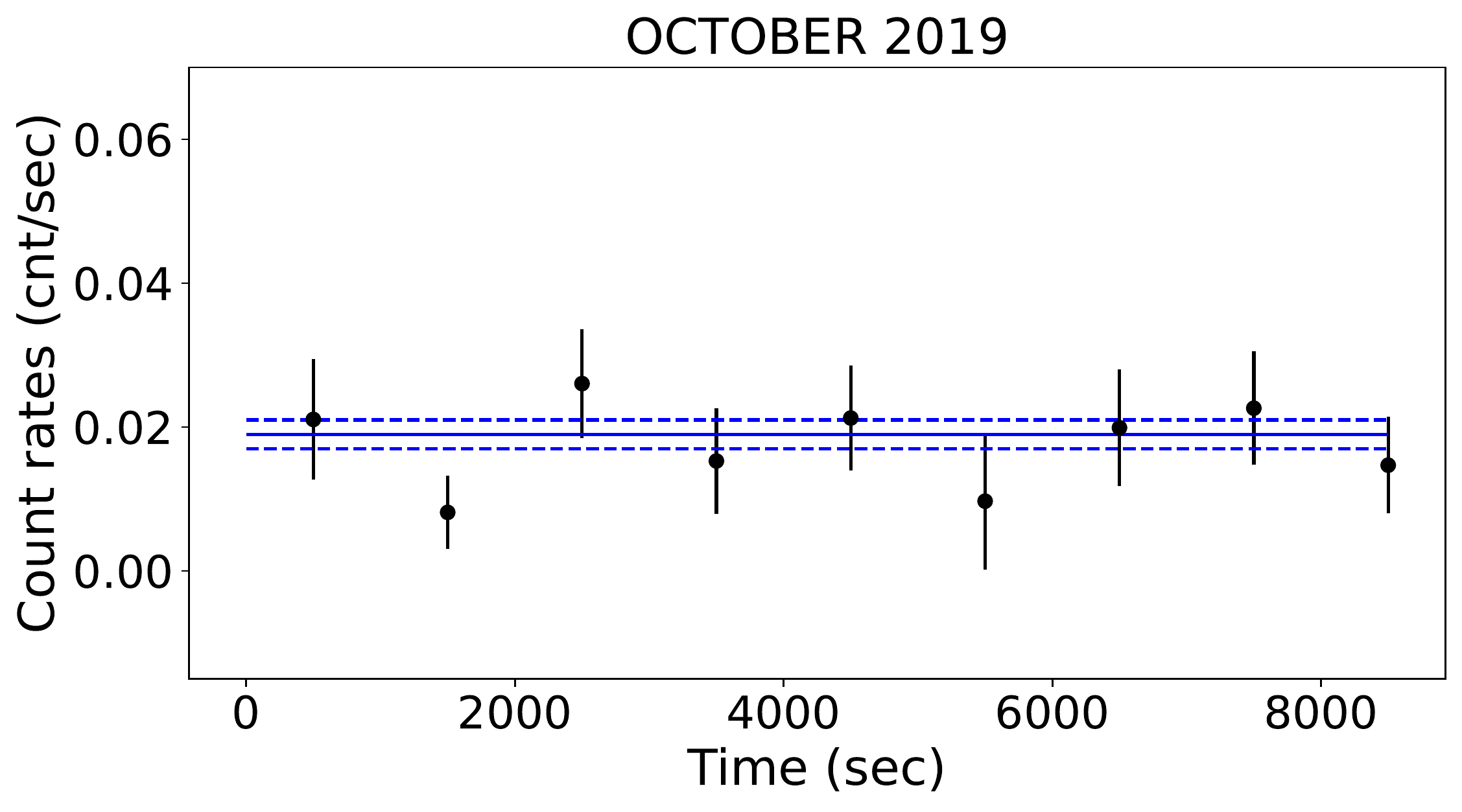}}\\
\subfloat{\includegraphics[width=0.5\textwidth]{ 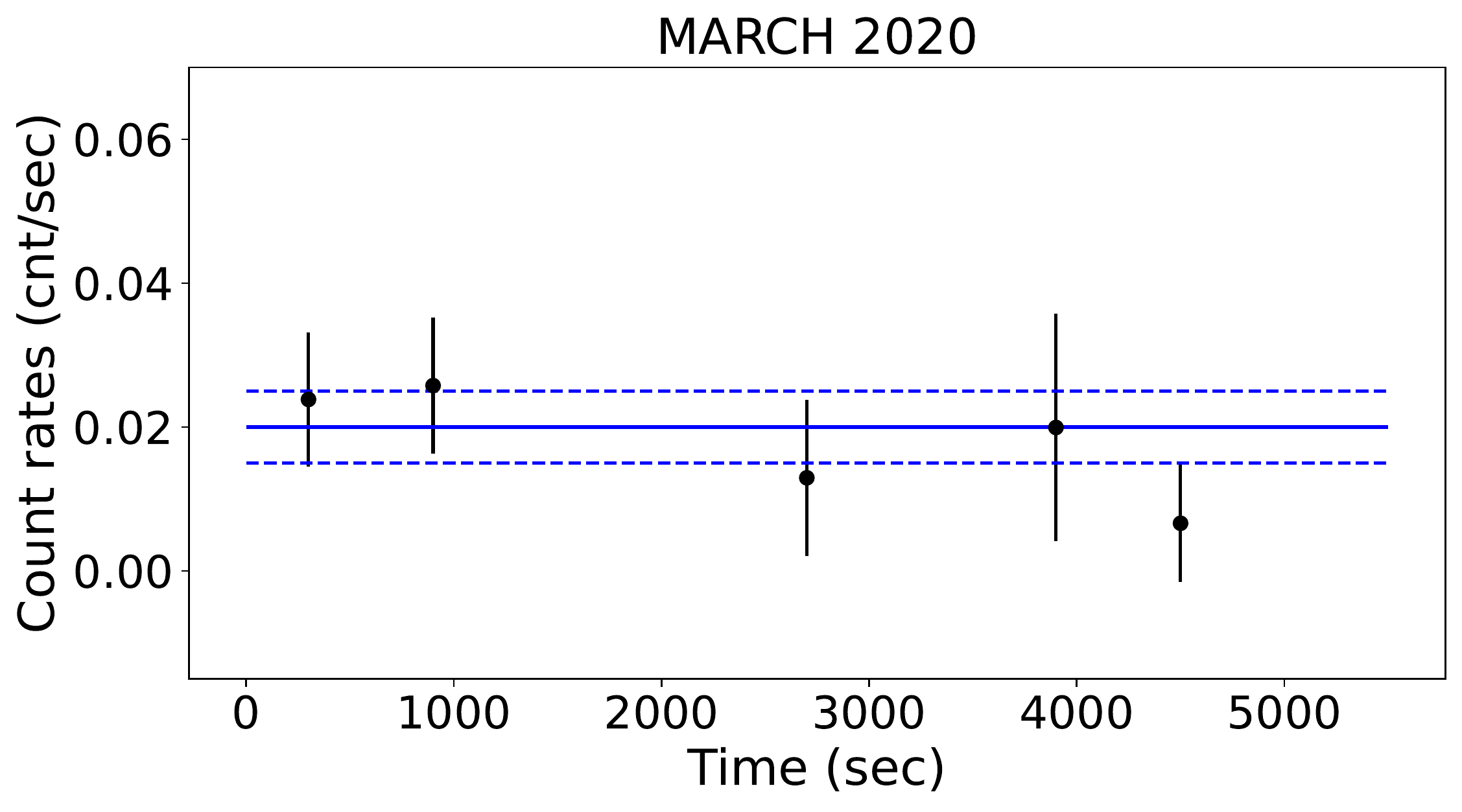}}\\
\caption{EPIC/pn background-subtracted lightcurves for all observations of Kepler 63. The blue lines are the mean count rate (solid lines) obtained from the source detection procedure, with the associated error (dashed lines).}
\label{fig:lc}
\end{figure*}

\ref{fig:lc} shows the background-subtracted lightcurves of Kepler-63. In these plots, the solid blue lines represent the time-averaged count rates calculated with the source detection procedure, and its uncertainties (dashed lines). These values are reported in \ref{tab:obs_log}. 

We checked for short-term variability in each ligthcurve, employing the software \texttt{R} and its package \texttt{changepoint} \citep{JSSv058i03}. This package allows to look for changes of mean and variance in time series data. The application of this method did not yield any significant short-term variability in our observations, i.e. no flare-like events occurred during the X-ray monitoring of Kepler-63.

\subsection{EPIC/pn spectra}
\label{sec:spectra_single}
The spectra of Kepler-63 were extracted for the whole energy band of \textit{XMM-Newton}, i.e. $0.2-12.0$~keV, and grouped with a minimum of counts per bin equal to 15, with the exception of the observation from March 2020. For this last observation, a grouping size of minimum 15 counts per bin would result in very few spectral bins, as
it is the one with the lowest photon statistics, and with a short exposure time.  Therefore, only the spectrum of the observation of March 2020 was grouped choosing 5 as minimum counts per bin. However, even with the reduced number of counts per bin, the spectrum of this observation has too few bins to be properly fitted and, thus, it was excluded from the analysis. 

Each of the other four spectra was analysed with the standard spectral software \texttt{xspec} (version 12.10; \citealt{1996ASPC..101...17A}) and was fitted with the simplest thermal model, i.e. a one-temperature (1-T) APEC model, 
considering the energy range $0.3-2.0$~keV \footnote{Beyond $2.0$~keV the spectra are extremely noisy and the high uncertainties would compromise the goodness of the fit.}.
The abundances $Z$ are kept frozen during the fitting procedure, on a value equal to $0.3Z_\odot$, which is typical for the corona of solar-like stars \citep{2007ApJ...660.1462M}. Tests for other values of abundances are explained in \ref{sec:abund}.
The photoelectric absorption is not considered in the spectral model: although Kepler-63 is at a distance for which this effect plays a relevant role, i.e. at $200$~pc, below $0.3$~keV the uncertainties on the counts are too high to make the spectrum physically reliable. 

\begin{table*}
\centering
\caption{Best-fitting parameters for the X-ray spectra of Kepler-63. The spectral model used for all spectra is the 1-T APEC model, with the exceptions of the observations of September 2014, August 2019 and the simultaneous fitting of all observations obtained in 2019: in these latter cases a 2-T APEC model was used as well. For both spectral models the metal abundances are kept frozen on the value $0.3Z_\odot$. The associated uncertainties are obtained with the \texttt{error} command of \texttt{xspec}. In the last column, $\overline{\chi}^2$ stands for the reduced $\chi^2$ and d.o.f are the degrees of freedom.}
\begin{minipage}{\textwidth}
\resizebox{\textwidth}{!}{
\begin{tabular}{ccccccccc}
\hline
Date       & Model & $kT_1$          & $kT_2$              & $logEM_1$         &$logEM_2$   & $kT_{\rm av}$ &  $F_{\rm X}$                            & $\overline{\chi}^2$ (d.o.f) \\

           &       & [keV]             & [keV]               & $[\rm cm^{-3}]$ &  $[\rm cm^{-3}]$ & [keV] &$10^{-14} \ \rm [erg/cm^2/s]$      &                     \\
           &       &                &                      &                   &     &      &  $0.3-2.$~keV                    &                     \\
           \hline \hline
2014-09-28 & 	1-T APEC&	$0.60 \pm	0.12	$&	$\smallsetminus$ 		&$	51.92	\pm	0.13	$&	$\smallsetminus$ 	&	$\smallsetminus$ 		&$	2.27 \pm	0.16	$&	0.74 (27)	\\
	&2-T APEC &	$0.50 \pm 	0.13	$&	$1.08 \pm	0.47	$&$	51.84	\pm	0.41	$&$	51.45	\pm	0.80	$& $0.67 \pm 0.18$ &           $	2.43 \pm	0.22	$&	0.64 (25)	\\
2019-05-05&	1-T APEC&	$0.79 \pm	0.29	$&	$\smallsetminus$ 		&$	51.83	\pm	0.34	$&	$\smallsetminus$ 	&	$\smallsetminus$ 		&$	1.91 \pm	0.35	$&	0.65 (11)	\\
2019-08-21&	1-T APEC&	$0.82 \pm	0.18	$&	$\smallsetminus$ 		&$	51.80	\pm	0.22	$&	$\smallsetminus$ 	&	$\smallsetminus$ 		&$	1.75 \pm	0.26	$&	1.39 (13)	\\
	&2-T APEC&	$0.27 \pm	0.09	$&	$1.04 \pm 	0.41	$&$	51.79	\pm	0.47	$&$	51.66	\pm	0.49	$& $0.61 \pm 0.18$ &   $	2.29 \pm	0.28	$&	0.43 (10)	\\
2019-10-31&	1-T APEC&	$0.74 \pm	0.20	$&	$\smallsetminus$ 		&$	51.92	\pm	0.30	$&	$\smallsetminus$ 	&	$\smallsetminus$ 		&$	2.38 \pm 	0.38	$&	0.81 (7)	\\
\hline \hline
Simultaneous fit (2019) &2-T APEC & $0.31 \pm 0.16$ & $0.94 \pm 0.23$ & $51.59 \pm 0.63$ & $51.72 \pm 0.38$ &$0.67 \pm 0.15$ & $1.93 \pm 0.19$  & $0.75$ ($32$) \\
\hline
\end{tabular}
}
\end{minipage}
\label{tab:1T}
\end{table*}
The best-fitting values resulting from this procedure are reported in \ref{tab:1T}. The uncertainties of each best-fitting parameter are obtained with the \texttt{error} command of \texttt{xspec}, that calculates the confidence interval for the model parameters within a confidence level of $90\%$.  The spectra of each observation, with the spectral model overplotted, are shown in \ref{fig:fit_spec}.

The observation from September 2014 and the one from August 2019 show sufficiently high statistics to be fitted with a 2-T APEC model. For these two spectra the best-fitting parameters for the 2-T model are also shown in \ref{tab:1T}. The abundances are again frozen to the value of $0.3Z_\odot$. 

\subsubsection{Metal abundances}
\label{sec:abund}
We carried out tests with the aim of constraining the coronal abundances for each observation. To this end, we adopted a 1-T APEC model in which the metal abundances were left free to vary during the fitting procedure. However, because of the low photon statistics that each spectrum shows, it was not possible to have a statistically significant fit. Only the fit of the archival observation yielded a constrained value for $Z$, equal to $0.24 \pm 0.07$, with a reduced $\chi^2$ of $1.25$ (with 25 d.o.f). 

We additionally noticed that when considering a range of different values of metal abundances, each of them kept frozen in the fitting procedure, the statistics of the fit did not show significant differences. Thus, with the quality of the available data, we can not constrain the coronal metal abundances. 

For the further analysis, we kept the metal abundances always frozen to $0.3Z_\odot$, as mentioned above. 
\subsubsection{Simultaneous fitting}
\label{sec:spectra}
We performed a simultaneous fit of all usable observations obtained in 2019 (three observations in total), in order to increase the signal-to-noise ratio. This allowed us to use a 2-T APEC model. The 2-T fit enables (1) a more detailed comparison to the observation from September 2014 and (2) a more meaningful investigation of Kepler-63 in the framework of the study "The Sun as an X-ray star" (see \ref{sec:models}).

We carried out the simultaneous fit as described in \ref{sec:spectra_single}. The results are shown in the bottom right panel of \ref{fig:fit_spec} and reported in \ref{tab:1T}.

\section{Modelling the corona of Kepler-63 in terms of solar magnetic structures}
\label{sec:models} 
Here, we present our technique for indirectly identifying magnetic structures on the corona of Kepler-63. As explained in \ref{sec:intro}, the method is based on the study "The Sun as an X-ray star" and was previously applied to the old binary star HD81809 \citep{2008A&A...490.1121F, 2017A&A...605A..19O} and the young solar-like star $\epsilon$~Eridani (\citealt{2020A&A...636A..49C}).

In \ref{sec:sim_summary}, a summary of the study "The Sun as an X-ray star" is presented, followed by the description of its application to the case of Kepler-63 (\ref{sec:sim}).

\subsection{The solar emission measure distributions}
\label{sec:sim_summary}
In the study "The Sun as an X-ray star", different types of magnetic structures, observed with the Soft X-ray Telescope (SXT) on board the solar satellite \textit{Yohkoh} on the corona of the Sun during the 22nd solar activity cycle, were spatially and temporally analysed. The three kinds of structures that were identified in the X-ray images of the Sun are quiet regions (also termed background corona, BKCs), active regions (ARs) and cores of active regions (COs). For each of these structures, a distribution of emission measures as a function of the temperature (EMD) was derived from the observations.  In this way, for each phase of the solar cycle it was possible to identify  the contribution of each type of magnetic structure to the total EMD. 

The \textit{Yohkoh} satellite  provided also several images of solar flares on the corona, and eight flares were analysed in time, from their rise throughout their decay \citep{2001ApJ...557..906R}. These flares range from Class C (relatively low energy flares) to Class X (the most energetic ones observed on the Sun), and the corresponding flare EMDs were retrieved. Moreover, \textit{Yohkoh} collected observations of flares with different filters, hard filters sensitive to plasma temperatures around and above $10^7$~K and soft filters for lower temperatures. The vast majority of the flares were observed with a pair of hard filters only; a few of them were observed also with a pair of soft filters, thus allowing a better reconstruction of the EMD at temperatures as low as $3$~MK. One of these flares (a Class M flare, namely a flare of mid energy) was observed with two pairs of filters (soft and hard) and analysed by  \cite{2001ApJ...557..906R}.
It was also possible to trace the temporal evolution of one AR and its corresponding CO from the rise to the decay (see Fig. 2 of \citealt{2004A&A...424..677O}). 

In the final paper of "The Sun as an X-ray star" \citep{2001ApJ...560..499O}, for each observed type of solar EMD, i.e. that of BKCs, ARs and COs, a corresponding synthetic spectrum was extracted, using the response matrix of X-ray satellites that can observe stars. This way, \cite{2001ApJ...560..499O} simulated solar coronal observations as if the Sun was observed by non-solar X-ray instruments.

\begin{figure}[!htbp]
\includegraphics[width=0.5\textwidth]{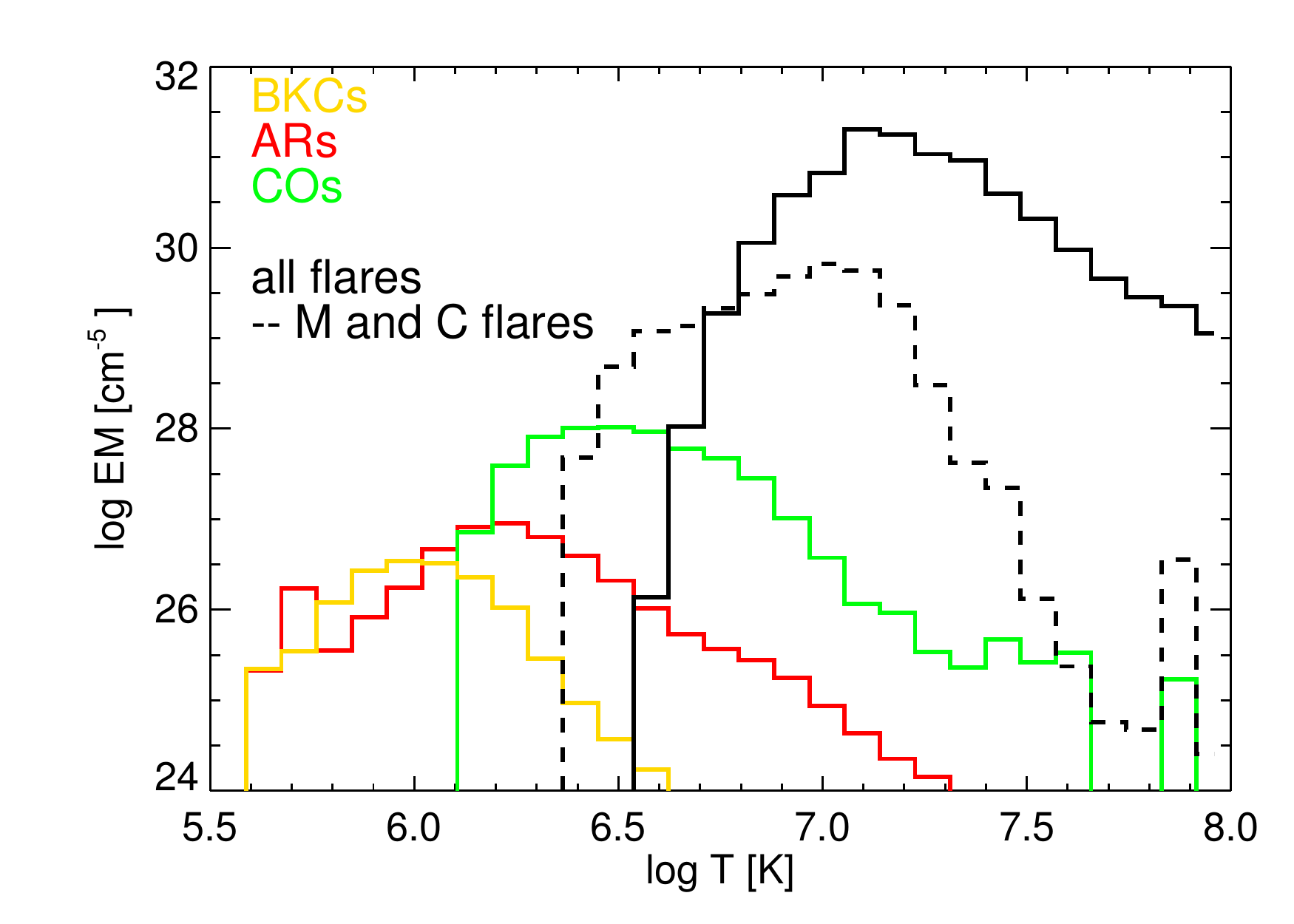}
\caption{Average EMDs per unit surface area for each type of magnetic structures observed with \textit{Yohkoh}. The black lines are the two flaring EMDs tested within this work: a time-averaged distribution derived from a sample of eight flares from Class C to X observed with a pair of hard filters with \textit{Yohkoh} (solid black line; \ref{sec:mod1}) and a time-averaged distribution derived from the flares of Class C and M in the sample (dotted black line; \ref{sec:mod2}), the M flares observed also with a pair of soft filters.}
\label{fig:all_st}
\end{figure}

\begin{figure*}[!htbp]
\centering
\includegraphics[width=\textwidth]{ 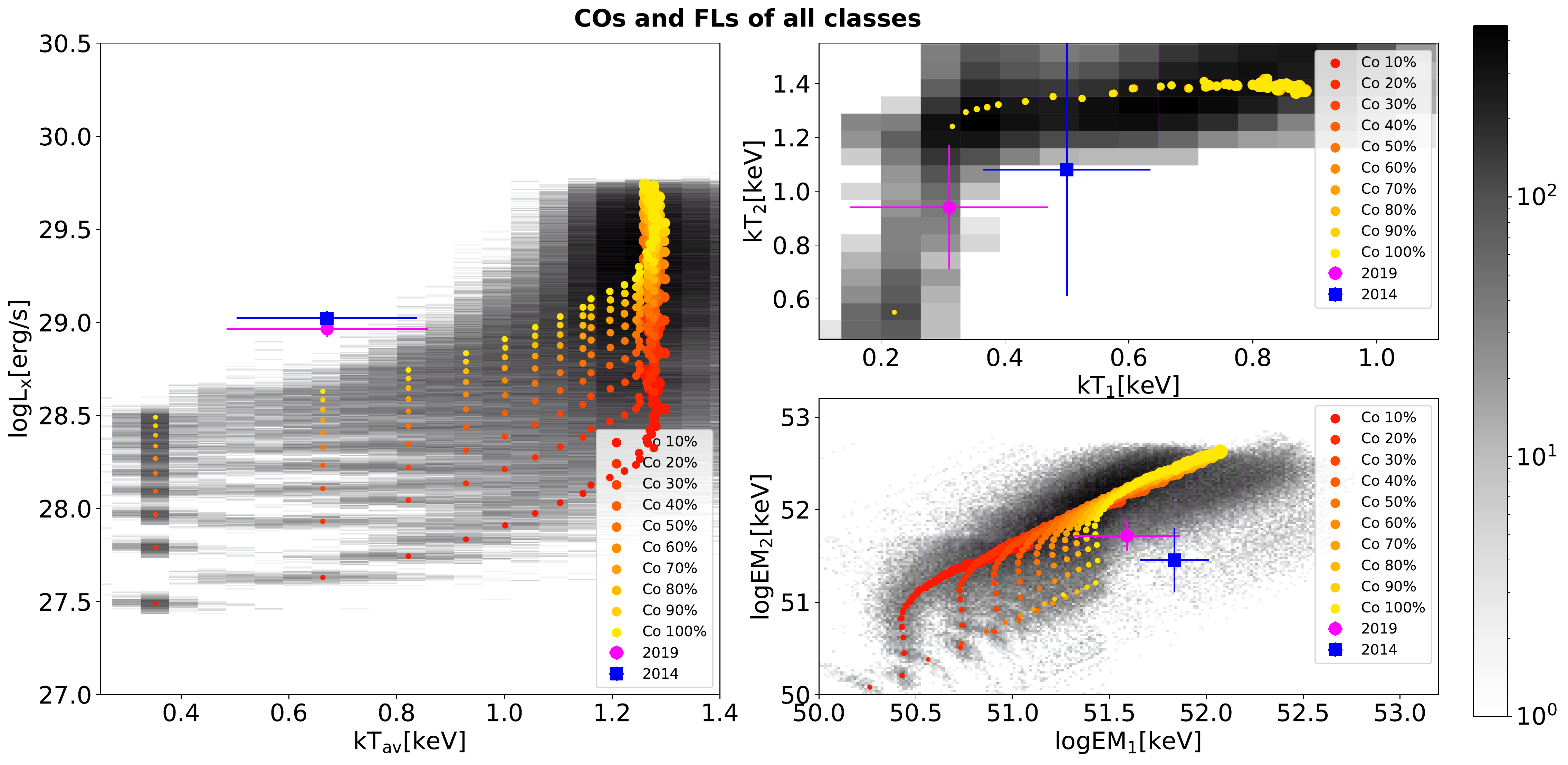}
\caption{X-ray luminosity as a function of emission measure weighted thermal energy $kT_{\rm av}$ (left panel) and the two temperatures and emission measures (right panel). The circles represent the synthetic spectra extracted from EMDs that are composed of time-averaged COs varying from a coverage of $10$ to $100\%$ of the total coronal surface (change of color shade) and by FLs, from Class C to X time-averaged over their evolution, that vary from $0$ to $4\%$ of the area of the COs (change of symbol size). The best-fitting parameters of the observations of Kepler-63 are overplotted with the blue square (2014) and the magenta circle (2019).In the background, the occurrences of each best-fitting parameter, that is simulated 1000 times for each combination of magnetic structures by introducing Poisson noise, are shown with the corresponding color bar in log scale.}
\label{fig:sim1}
\end{figure*}

\begin{figure*}[!htbp]
\centering
\includegraphics[width=\textwidth]{ 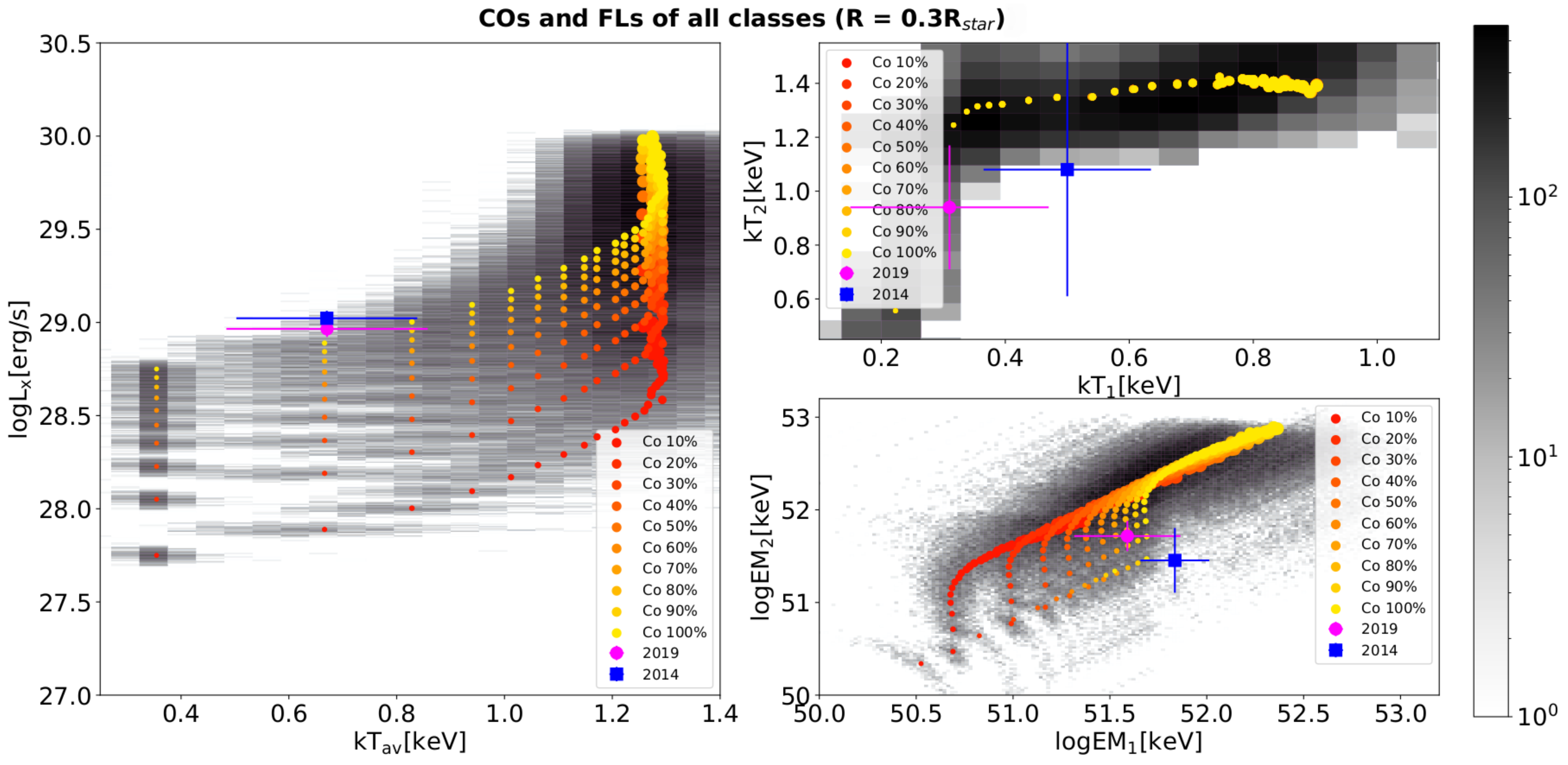}
\caption{X-ray luminosity as a function of emission measure weighted thermal energy $kT_{\rm av}$ (left panel) and the $kT$s and $EM$s (right panel). The EMDs used for these grids assume the same magnetic structures and the same percentage coverage as in \ref{fig:sim1}, with the difference that each EMD is scaled to a larger extension of the corona of Kepler-63, i.e. the stellar radius is set to be $30\%$ larger. The coding of the plots follows \ref{fig:sim1}.}
\label{fig:sim2}
\end{figure*}

\subsection{The case of Kepler-63}
\label{sec:sim}
With the aim of investigating whether and how the corona of young active stars can be described by magnetic structures similar to those observed in the solar corona, we applied the method presented in \ref{sec:sim_summary} to Kepler-63. The Sun is, indeed, the only star on which magnetic structures can be spatially and temporally resolved with current technology, and thus the corona of the Sun is the only template that can be used for such investigations.

For applying the above study, we considered that at least two spectral components are needed in order to achieve a basic description of the temperature structure of the corona. However, the individual observations of the 2019 campaign are too poor for a 2-T fit, and thus we have fitted the three spectra jointly (see \ref{sec:spectra}). We, therefore, have two epochs with spectral fits for Kepler-63 available: one representing the archival observation (from 2014) and one representing the average corona observed in 2019. 
 
\subsubsection{Choice of the appropriate magnetic structures}
\label{sec:choice_ms}
In the first step we aimed at finding the appropriate types of magnetic structures whose EMDs, scaled to the coronal metal abundance (i.e. $0.3Z_\odot$) and the surface area of Kepler-63 ($R_\star=0.91R_\odot$; \citealt{2013ApJ...775...54S}), are suited to reproduce the X-ray luminosity ($L_{\rm X}$) and emission measure weighted thermal energies ($kT_{\rm av}$) observed during the \textit{XMM-Newton} monitoring campaign of the star. The $kT_{\rm av}$ is calculated as
\begin{equation}
kT_{\rm av} = \dfrac{ kT_1 \cdot EM_1 + kT_2 \cdot EM_2}{ EM_1+EM_2},
\label{eq:temp}
\end{equation}
where $kT_i$ and $EM_i$ ($i=1,2$) are the spectral components of the 2-T APEC model.

In \ref{fig:all_st} we show the EMDs derived from \textit{Yohkoh} data for the different types of coronal structures: BKCs (yellow line), ARs (red line) and COs (green line). The figure also shows two EMDs for flaring regions: one derived from a sample of flares from Class C to X observed with \textit{Yohkoh}/SXT using a pair of hard filters (\citealt{2001ApJ...557..906R}; solid black line) and the other one derived from flares of the sample from Class C to M (dashed black line), with the latter analysed also using a pair of soft filters in addition to the hard ones. 

From \ref{fig:all_st} we see that the EMD per unit surface area of the solar BKCs and ARs are much lower than the one of the COs and, thus, any percentage of these two structures would negligibly contribute to the total EMD. %(see Fig. 5 in \citealt{2020A&A...636A..49C}).
Covering part of the corona with  BKCs and/or ARs implies less space left available for other magnetic structures that have a higher surface brightness and would better reproduce the X-ray luminosity of Kepler-63, which is at least two orders of magnitude higher than that of the Sun ($\log L_{\rm X \ \rm Kep63}  \rm \ [erg/s] \sim 29$ versus $\log L_{\rm X \ \odot} \rm \ [erg/s] \sim 27.3 $; \citealt{2001ApJ...560..499O}). We interpret this as a clear indication that Kepler-63 requires a high surface coverage with COs, so that the contribution of BKCs and ARs to the X-ray emission is negligible. Thus, we do not consider the contribution of BKCs and ARs in the following. 

For the case of $\epsilon$~Eridani, \citet{2020A&A...636A..49C} used the EMD obtained from the CO observed in 1996 by \citet{2004A&A...424..677O}, time-averaged during its evolution (solid green line in \ref{fig:all_st}), together with the contribution of the EMDs of FLs. These EMDs of COs and FLs  were able to reproduce the X-ray luminosity of $\epsilon$~Eridani ($\log L_{\rm X \ \epsilon Eri} \rm \ [erg/s] \sim 28 $). 
Due to the higher X-ray luminosity of Kepler-63 ($\log L_{\rm X \ \rm Kep63}  \rm \ [erg/s] \sim 29$), it is necessary to increase the percentage of surface coverage of COs and FLs in order to reach the observed X-ray emission level of the star. 
We, therefore, started to test the same scenario adopted for the case of $\epsilon$~Eridani, where the synthetic corona is covered with time-averaged COs and FLs, albeit assuming a greater percentage of them than in the case of $\epsilon$~Eridani. %This increases the X-ray luminosity and average temperature of the simulated corona. 
For the flaring contribution, we tested two different FL EMDs as we describe in \ref{sec:mod1} and \ref{sec:mod2}. These two distributions are plotted in \ref{fig:all_st} as solid and dotted black lines respectively.   

We took into account that the different classes of flares have different occurrence rates and, therefore, the flare filling factor depends on the flare energy (here approximated by peak flux, alias GOES class). To determine the average contribution of the different flare classes to the total EMD we assumed that the differential frequency of the peak flux ($F_{\rm peak}$) of solar flares is described as $dN(F_{\rm peak}) \propto F_{\rm peak}^{-\alpha}dF_{\rm peak}$ with index $\alpha = 1.57$. This is the value calculated by \cite{Aschwanden_2002} from \textit{Yohkoh}  observations of solar flares.  The flare flux frequency distribution (FFD) is normalized to the peak flux of the flare with the lowest flux (a Class C flare in the case of the sample of flares considered here). The flare filling factors we derive with our modelling thus apply to a Class C5.8 flare, and filling factors for flares with other peak fluxes can be obtained by scaling from the power-law (see \ref{sec:flares}). We note that the normalization of the FFD is degenerate with the filling factor of flares. If the normalization was made on a less energetic flare the absolute values for the flare frequencies of the more energetic events would be lower, and a higher filling factor would be required to compensate for this.

After the appropriate classes of magnetic structures have been defined (here, COs and FLs), a grid of EMDs was constructed where each grid point is distinguished from the others by a different fractional surface coverage with each class of magnetic structures.  

\subsubsection{Synthesis of the X-ray spectra}
We extracted a synthetic\textit{ XMM-Newton} spectrum for each grid point, making use of the EPIC/pn response matrix and taking into account the distance of Kepler-63. In order to estimate uncertainties, 1000 spectra were generated for each grid point, and Poisson statistics were introduced in the extraction procedure for the synthesis of the spectra. Each of these 1000 spectra was then fitted in \texttt{xspec} with a 2-T APEC model, and with metal abundances frozen to $0.3Z_\odot$, i.e. the same spectral model used to represent the observed spectra of Kepler-63.
In this way, each grid point, i.e. each combination of magnetic structures, is not unequivocally represented by only two couples of $kT$s and $EM$s for the cold and hot components, but rather by a set of 1000 values for each of the four fit parameters drawn from a Poisson distribution. The four parameters (two $kT$s and two $EM$s) for each point of the final grid are calculated as the mean of the 1000 values obtained from the spectral analysis.

Furthermore, from analyzing the synthetic spectra we obtained the X-ray luminosity as a function of the EM weighted coronal temperature, calculated as in Eq. \ref{eq:temp}, and we performed a first visual comparison of these values with those calculated from the spectra of Kepler-63, in order to verify if the chosen combination of magnetic structures can reproduce the observations.  
The left panel of \ref{fig:sim1} is an example of $L_{\rm X}$ as a function of $kT_{\rm av}$, where the grid and the observations are compared. We discuss the details of such comparisons for different grids in \ref{sec:mod1} and \ref{sec:mod2}.

As a final step, we performed a quantitative match between the temperatures and the emission measures, obtained from the analysis of the synthetic spectra with those retrieved from the observations, according to the approach described by  \citet{2020A&A...636A..49C} and summarized in the subsequent section.

\subsubsection{Selection criterion}
\label{sec:crit}
The match between synthetic and observed X-ray spectra of Kepler-63 was performed by using a set of four equations, one for each of the four best-fitting parameters ($kT_i$ and $EM_i$, with $i=1,2$), following \citet{2020A&A...636A..49C}:
\begin{equation}
P_i ^{\rm obs} - \Delta P_i ^{\rm obs} \cdot \sigma \leq P_{i,j,k} ^{\rm syn} \geq P_i ^{\rm obs}  + \Delta P_i ^{\rm obs} \cdot \sigma,
\label{eq:crit}
\end{equation}
where $P_i ^{\rm obs}$ are the parameters obtained from the spectral analysis of the observations of Kepler-63, and $\Delta P_i ^{\rm obs}$ are the corresponding errors. $P_{i,j,k} ^{\rm syn}$ are the best-fitting parameters of the synthetic spectra. The index $j$ denotes the 1000 spectra, randomized with a Poisson distribution, generated for each $k$-th grid point \footnote{The index $k$ changes according to the number of combinations of
magnetic structures present in a given grid. In this work, the final grids (i.e. colored circles in \ref{fig:sim1}, \ref{fig:sim2} and \ref{fig:sim3}) were composed from $\sim 400$ to $\sim 1500$ points.}. 

In Eq. \ref{eq:crit}, the parameter $\sigma$ is the same for all four equations and determines the global confidence range of the match between observed and synthetic values. Thus, the criterion finds the set of best-fitting parameters for each 1000 sets ($j$) of spectra among all grid-points ($k$) within the smallest $\sigma$. At the end of the selection procedure there are in total 1000 best matches between one observation and the synthetic spectra, that correspond to a range of best-fitting combinations of young COs and FLs. From these 1000 representations, the unique best-matching combination of young COs and FLs was found from the median of the $j$ selected values, associating as error the $10\%$ and $90\%$ quantile of these values.

\begin{figure*}[!htbp]
\subfloat{\includegraphics[width=0.5\textwidth]{ 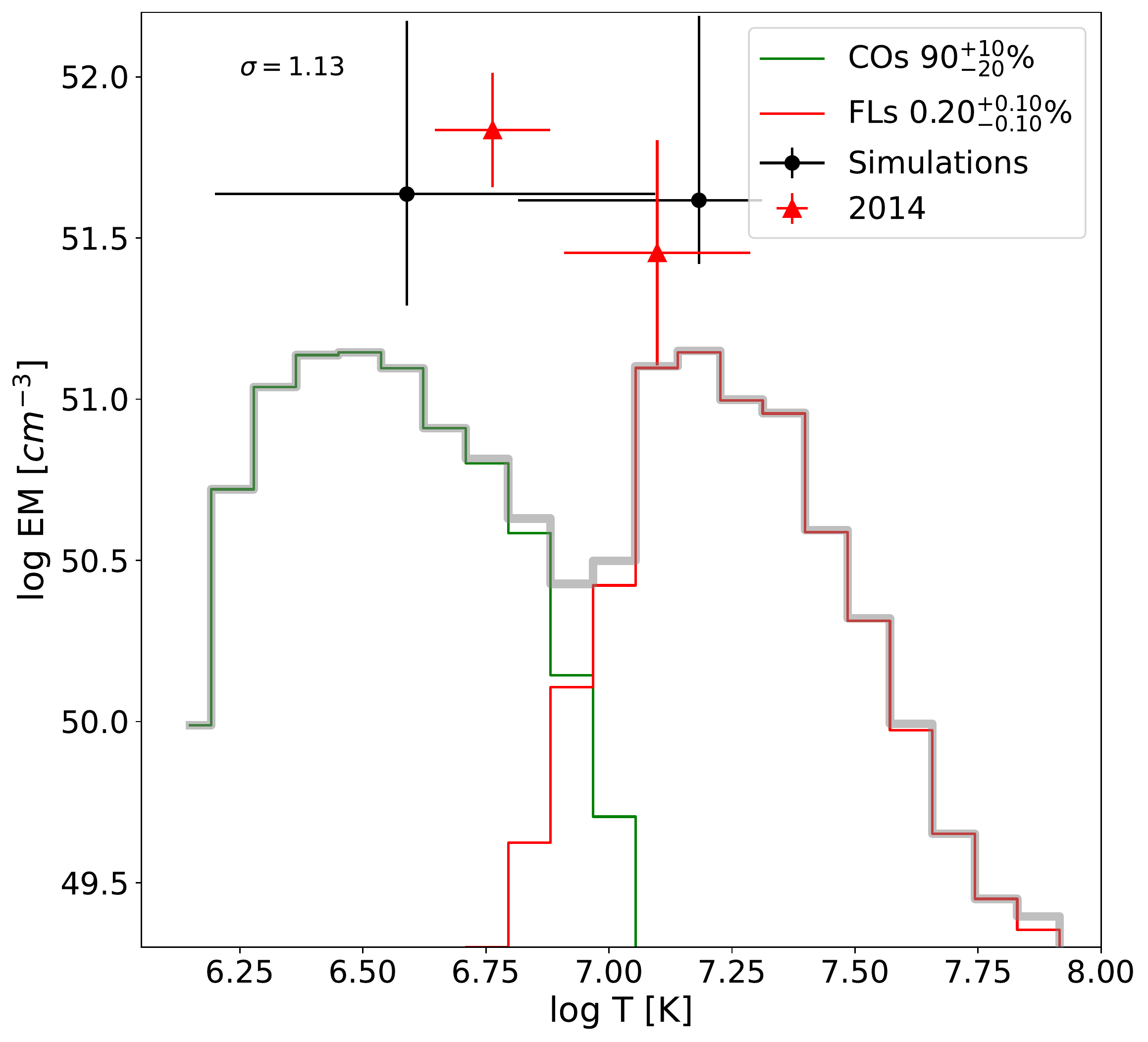}}
\subfloat{\includegraphics[width=0.5\textwidth]{ 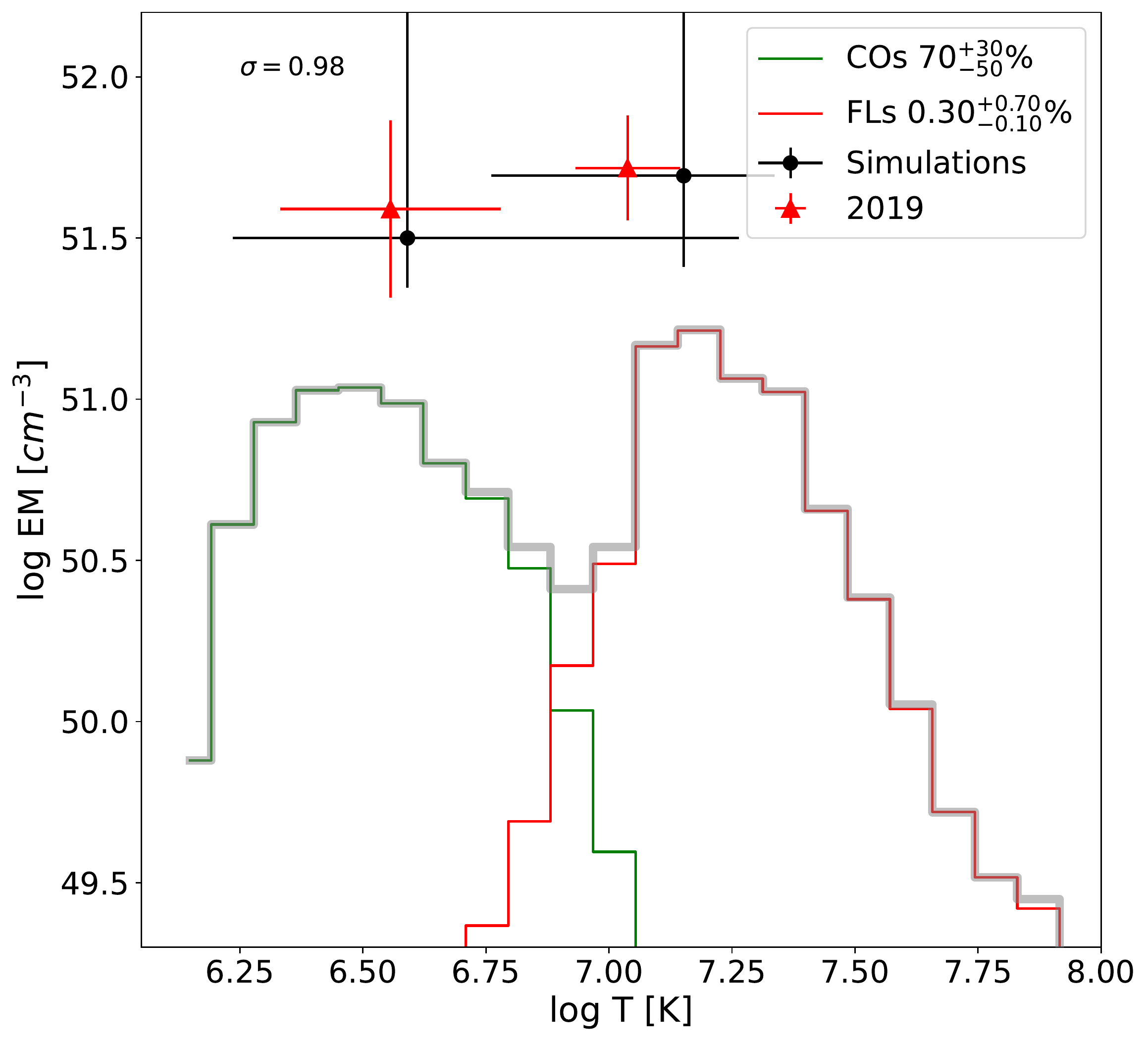}}
\caption{EMDs constructed from solar magnetic regions that best match the observations of Kepler-63 (on the left: 2014; on the right: 2019).  The contributions to the total EMD (grey line) are from time-averaged COs (green line) and FLs (red line) from Class C to X time-averaged over their evolution. The confidence level $\sigma$ within which the models are selected is given on the upper left corner of each plot. The EMDs are scaled to a larger size of the corona of Kepler-63, i.e. $+30\%$ than its stellar radius. The red squares are the best-fitting parameters of the X-ray spectra of Kepler-63. The black circles are the best-fitting parameters retrieved from the best-matching synthetic spectra. The filling factor for the best-matching combination of magnetic structures is given in the legend. }
\label{fig:bestMod1}
\end{figure*}

\begin{figure}
\includegraphics[width=0.5\textwidth]{ 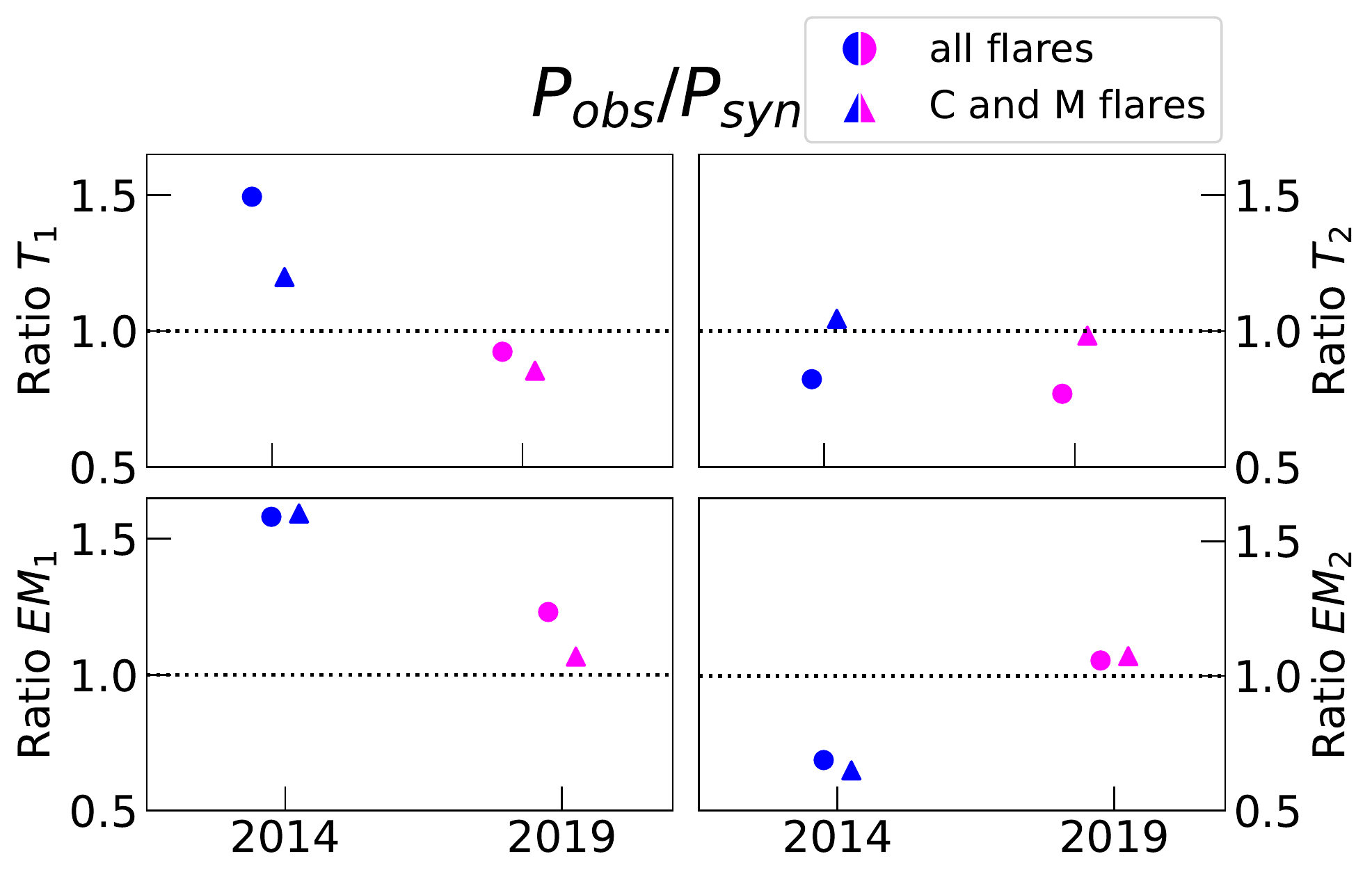}
\begin{minipage}{0.5\textwidth}
\caption[BB]{Ratios between the best-fitting parameters of the spectra of Kepler-63 and the $kT$ and $EM$ values of the best-matching synthetic spectra\footnote{The error bars are not taken into account as the calculated uncertainties are greater than the corresponding ratios.}. The blue symbols are the ratios obtained from the comparison with the observation of 2014, whereas the magenta ones are obtained for the 2019 observations. The circles refer to those parameters calculated from the best-matching EMDs that have as flaring contribution the FLs from Class C to X time-averaged over their evolution (\ref{sec:mod1}). The magenta triangles are instead calculated from the best-matching EMDs that include flares of Class C and M time-averaged over their evolution. (\ref{sec:mod2})}
\label{fig:ratio}
\end{minipage}
\end{figure}

\begin{figure*}[!htbp]
\centering
\includegraphics[width=\textwidth]{ 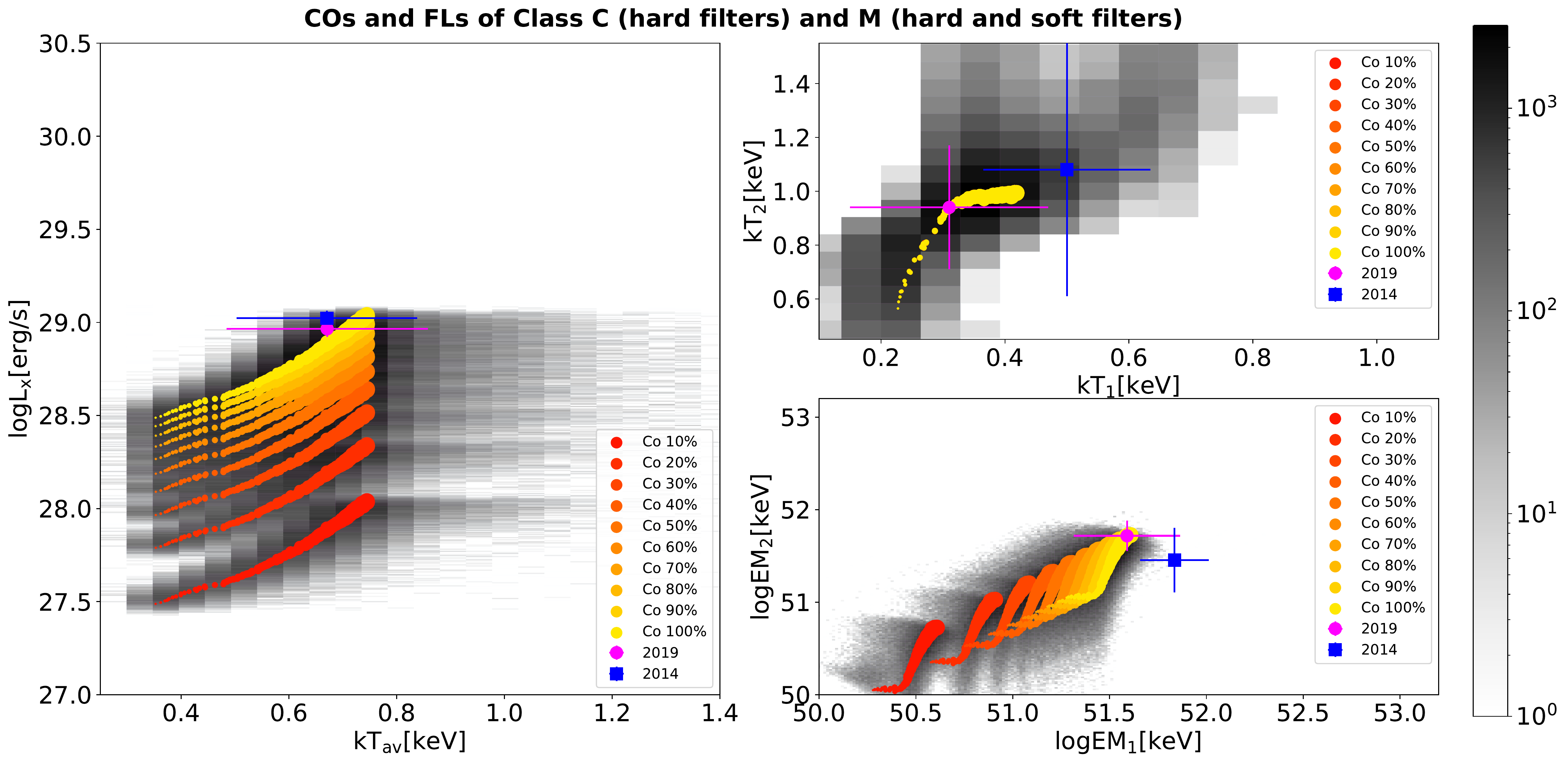}
\caption{X-ray luminosity as a function of emission measure weighted thermal energy $kT_{\rm av}$ (left panel) and the two temperatures and emission measures (right panel). The EMDs used for generating these grids are composed of time-averaged COs covering from $10$ to $100\%$ of the total corona and FLs of Class C and Class M, time-averaged over their evolution, that vary from $0$ to $15\%$ of the area of the COs. The Class M flares were also observed with \textit{Yohkoh} with a pair of soft filters in addition to the pair of hard filters (used for the analysis of the Class C flare). The coding of the plots follows \ref{fig:sim1}.}
\label{fig:sim3}
\end{figure*}

\begin{figure*}[!htbp]
\subfloat[]{\includegraphics[width=0.5\textwidth]{ 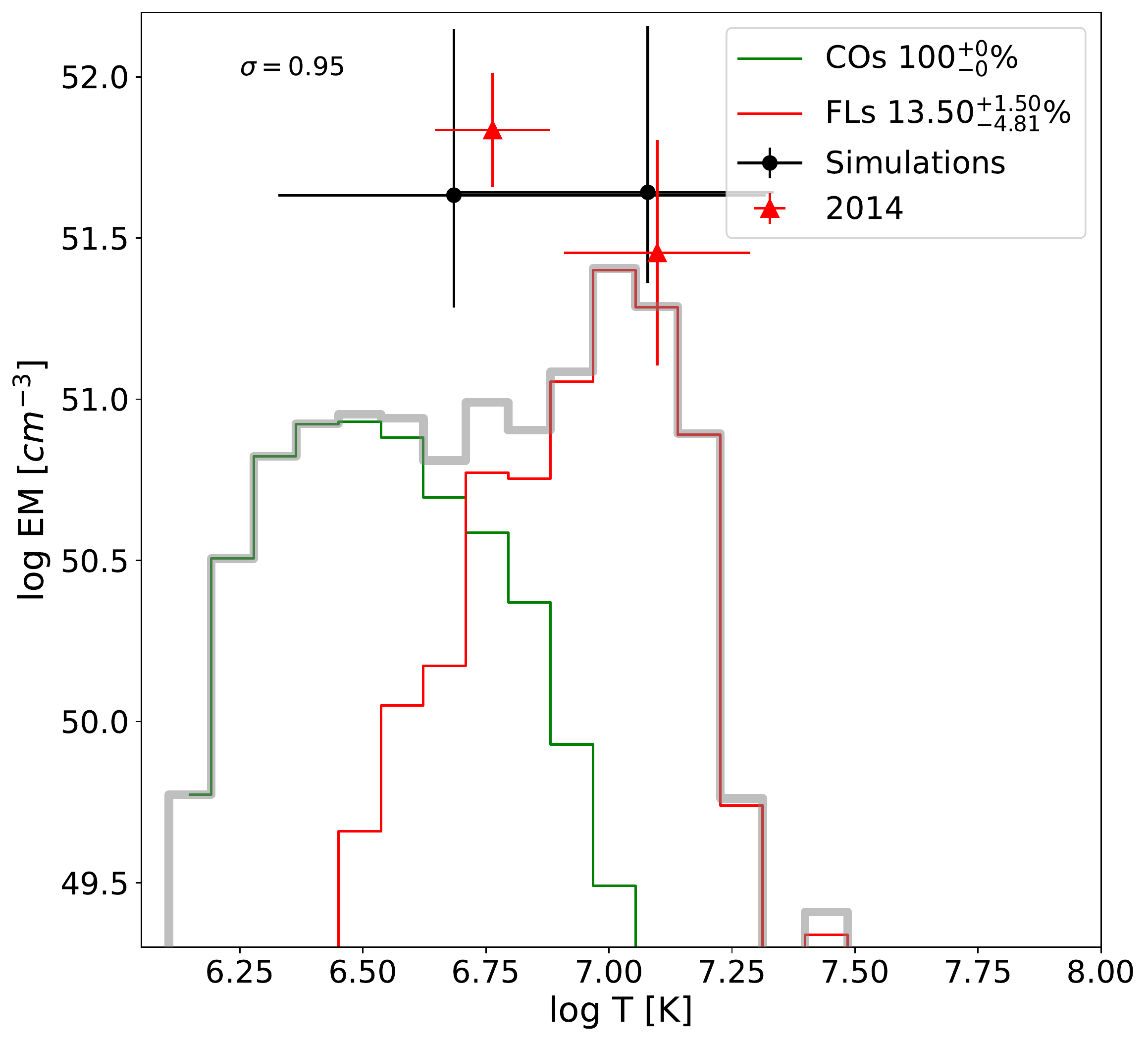}}
\subfloat[]{\includegraphics[width=0.5\textwidth]{ 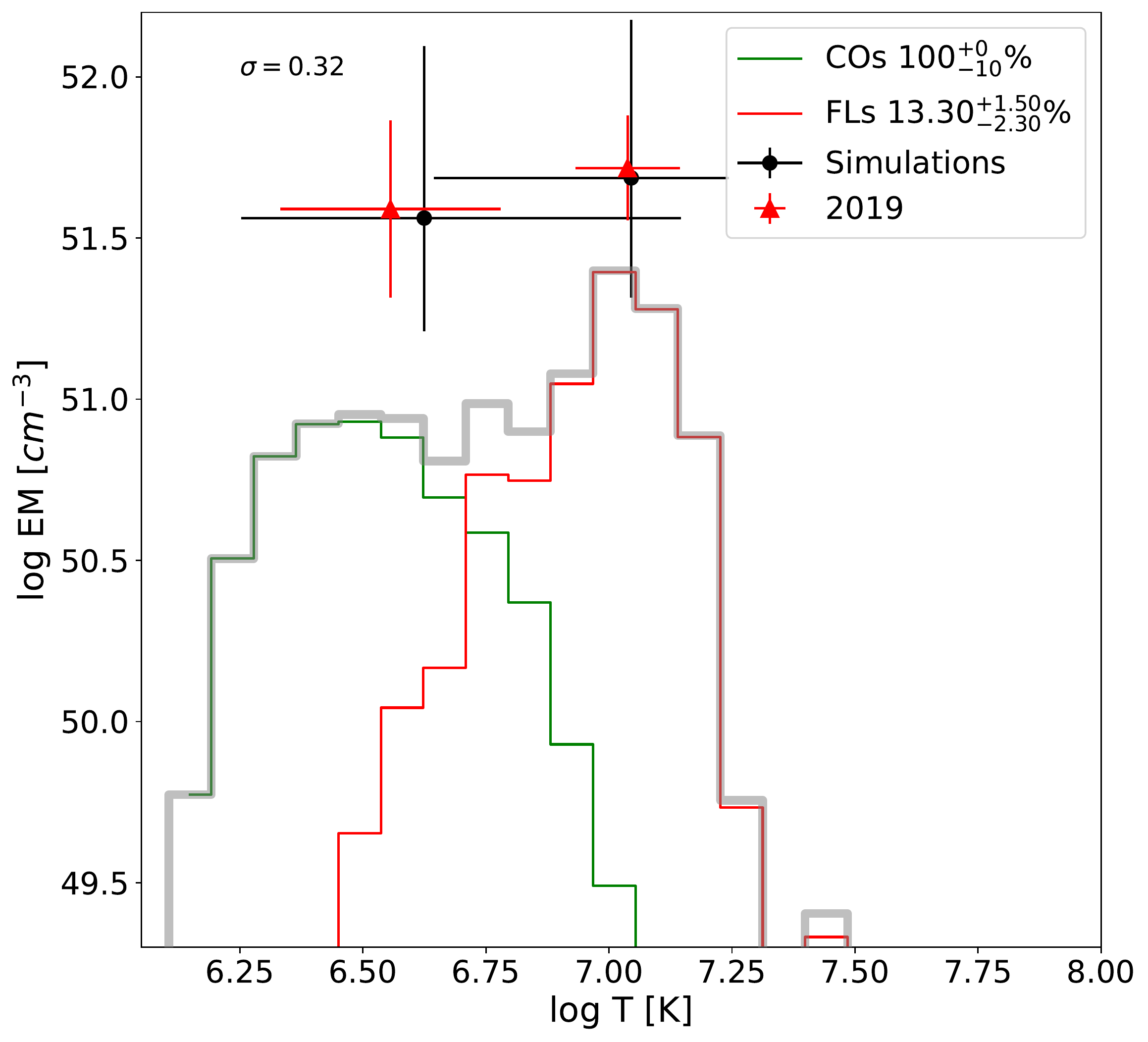}}\\
\caption{EMDs constructed from solar magnetic regions that best match the observations of Kepler-63.  The contributions to the total EMDs are from COs and FLs of Class C and Class M, time-averaged over their evolution. The confidence level $\sigma$ within which the models are selected is given on the upper left corner of each plot. The coding of the plots follows \ref{fig:bestMod1}.}
\label{fig:bestMod2}
\end{figure*}

\subsubsection{Match with the standard flaring Sun}
\label{sec:mod1}
The first grid of EMDs, and corresponding spectra, that we tested is composed of time-averaged COs, varying from $10\%$ to $100\%$ coverage of the corona, and by a distribution of FLs, time-averaged over their evolution, that vary from $0\%$ to $4\%$ of the COs. 
The FL distribution was computed by considering a sample of flares from Class C to Class X observed with \textit{Yohkoh}, from their rise to their decay phase (the average EMD per unit surface is shown in \ref{fig:all_st} as black solid line), and taking into account the power-law FFD as explained in \ref{sec:choice_ms}.

The plot of the X-ray luminosity as a function of the $kT_{\rm av}$ calculated from this grid is shown in \ref{fig:sim1} on the left: each of the circles represents one point of the synthetic grid, i.e. one specific combination of magnetic structures. The percentage of COs is designated with the color of the circles, whereas the percentage of FLs is represented by the size of the circles. It can be noticed that, as the coverage fraction of COs increases on the corona, the X-ray luminosity increases while the EM-weighted thermal energy is less affected. Similarly, the greater the percentage of FLs is, the higher the EM-weighted thermal energy results, whereas the variation of X-ray luminosity is moderate \footnote{This trend of X-ray luminosity and EM-weighted thermal energy, induced by magnetic structures, was already noticed for the Sun by \cite{2004A&A...424..677O}.}. 
This general behaviour of the grid is the same we found for $\epsilon$~Eridani \citep{2020A&A...636A..49C}.

The best-fit parameters obtained from the spectral analysis of Kepler-63 (see \ref{tab:1T}) are plotted with the blue and magenta symbols. The right panel in \ref{fig:sim1} shows the thermal energies and emission measures calculated from the grid and from the spectra of Kepler-63. The color and size code of the plotting symbols are the same as in the left panel. We can see that an increase of the percentage of COs corresponds to an increase of $EM_1$ and $EM_2$ only, whereas the FLs influence both thermal energies and the $EM_2$.

In the background of each plot, the number of occurrences for each pair of the best-fitting parameters, obtained from the 1000 simulations of each $k$-th grid point, are shown. The spread of these values provides an indication for the range of synthetic parameters compatible with the observations.  

In \ref{fig:sim1}, the $L_{\rm X}-kT_{\rm av}$ plot shows that the chosen combination of magnetic structures can reproduce the EM-weighted thermal energy of Kepler-63, but the observed X-ray luminosity is not reached by the grid, i.e. even a full surface coverage ($100\%$) with COs and FLs does not produce enough $L_{\rm X}$ at the observed values of $kT_{\rm av}$ to explain the brightness of the corona of Kepler-63.

We note that the simulations presented above do not consider any extension of the corona above the chromosphere. In other words, the corona is assumed to have a thickness close to zero. However, the observations of the Sun show that the corona expands significantly above the chromosphere, with coronal magnetic loops of ARs extending up to $20-30\%$ of the solar radius \citep{2007ApJ...661..532D}. It is, therefore, reasonable to consider an extension of the corona of Kepler-63 above the limb.

By assuming the same types of magnetic structures (time-averaged COs and FLs) with the same range of fractional coverage as before, we produced a grid of EMDs adopting a $30\%$ larger radius for the corona of Kepler-63 to account for its extension above the chromosphere. The left panel in \ref{fig:sim2} shows $L_{\rm X}$ as a function of $kT_{\rm av}$ calculated from this grid. The coding of the plot follows that of \ref{fig:sim1}. Here, the X-ray luminosity reproduces the observations of Kepler-63, although at the limit within the error bars. 
We, therefore, examined this model in more detail. 

First, by adopting the selection criterion of Eq. \ref{eq:crit}, we obtained the EMDs that best match the observations, shown in \ref{fig:bestMod1}. In these plots, the grey line is the total EMD that best matches the observations of  Kepler-63, and the green  and red lines are the EMDs of the COs and FLs that contribute to the total distribution at a given fraction coverage which is reported in the legend of the plots. The red and black symbols are the best-fitting parameters obtained from the spectral analysis of the actual spectra of Kepler-63 and the synthetic ones, respectively. 

In \ref{fig:ratio} the ratio between the best-fitting parameters of the spectra of Kepler-63 and those of the best-matching synthetic spectra are plotted as circles. It can be noticed that for the observations of 2019 (magenta circles) there is a better agreement with the selected models than for the observation of 2014 (blue circles) that shows larger discrepancies in particular in the ratios of the parameters of the first spectral component.

\subsubsection{Match with the modified flaring Sun}
\label{sec:mod2} 
To reduce the discrepancies seen in \ref{fig:sim2} and \ref{fig:bestMod1} between the best-fitting parameters of the models and those of the observations of Kepler-63, we need to change the components of the EMDs. As can be seen in \ref{fig:sim2} (right panel), a grid with lower $kT_2$ and $EM_2$ would better represent the data (especially for 2014). It is also clear in the plot that the COs reach the high X-ray luminosity of Kepler-63, and, therefore, this kind of structure is needed. We can thus modify only the FL component, and, as from \ref{fig:sim2} we see that we need smaller values of $kT_2$, a FL distribution composed of mid-energetic flares appears more suited.

Among the solar flaring observations acquired with \textit{Yohkoh} and presented by \cite{2001ApJ...557..906R}, there are two flares of Class M that were observed by the satellite with a pair of soft filters, in addition to the pair of hard filters that are most frequently used for \textit{Yohkoh} observations of flares. The soft filters yield an EMD that extends to lower temperatures, thus providing a more complete description of the EMD of flaring regions (see Reale et al. 2001, for more details). This is seen in \ref{fig:all_st} where the EMD  that includes the soft-filter observations (dotted black line) extends to lower temperatures than the time-averaged EMD of all (hard-filter) flares in the sample observed with \textit{Yohkoh}.

We, therefore, changed the FL distribution using the EMDs of solar flares of Class M, analysed with the two pairs of filters (soft and hard) and time-averaged across their evolution. Together with the Class M flares, we also considered the contribution of \cite{2001ApJ...557..906R}'s Class C flare to the EMD, as from the canonical flare frequency distribution (see \ref{sec:choice_ms}) the number of Class C flares is higher than that of Class M events, and therefore the former ones can not be excluded. 

In this new simulation the grid range of the coverage fraction of COs remains unchanged, that is varying from $10$ to $100\%$. The FLs are instead allowed to vary in a broader range of CO filling factor than before, namely $0-15\%$. This is motivated by the fact that the new FL component show an EMD per unit surface area smaller than the flares from Class C to X used in the previous grid (see the dotted black line versus the solid one in \ref{fig:all_st}). Moreover, the EMD of Class M flares covers a wider range of temperatures than the one of the time-averaged flares at different energies. This is due to the fact that the flaring distribution of Class M FLs was derived by using both soft and hard filters, providing a contribution to the EMD at lower temperatures.
 Thus, to reproduce the observed $kT_{\rm av}$, we needed to assume a higher percentage of FLs than in the previous case. 

Starting from these EMDs, we extracted the corresponding synthetic spectra and we analysed them as we have done previously, using a 2-T APEC model. Then, we proceeded with the selection criterion explained in \ref{sec:crit}. 

In \ref{fig:sim3}, the plot of $L_{\rm X}$ as a function of $kT_{\rm av}$ (on the left) and the plots of the $kT$s and $EM$s (on the right) are shown. The coding of these plots is the same as for \ref{fig:sim1}. Such a high percentage of FLs reproduces the X-ray luminosity without the need to account for the thickness of the corona of Kepler-63\footnote{We note that neglecting the thickness of the corona corresponds to deriving an upper limit to the surface coverage of COs and FLs; see discussion in \ref{sec:disc_ms}.}. In \ref{fig:bestMod2}, the EMDs that best match the observations are shown. In \ref{fig:ratio}, the ratios between the observed best-fitting parameters and the ones of the best-matching synthetic spectra for this test are shown as triangles. Here, it can be noticed that overall the ratios between the temperatures are closer to one than in the previous case, indicating that the replacement of the flaring EMD has led to a better agreement between the observations and synthetic spectra. On the other hand, the $EM$ values, and in particular those of the 2014 observation,  have not changed strongly with respect to the previous combination of magnetic structures.

\section{Discussion}
\label{sec:disc}
Kepler-63, a solar-like star with an age of only $\sim 200$~Myr, is the youngest star that so far was observed in X-rays with \textit{XMM-Newton} with the goal of searching for a coronal magnetic cycle. Our X-ray monitoring campaign consisted in four snapshots of its corona, distributed between 2019 and 2020, thus spanning 3/5 of the photospheric cycle ($P_{cycl}= 1.27$~yr; \citealt{2016ApJ...831...57E}).

One additional observation was available in the \textit{XMM-Newton} archive, dated back to 2014. \cite{2018MNRAS.477..808L} analysed this observation, using a 2-T APEC model, and they obtained an X-ray luminosity $\log L_{\rm X} \ [\rm erg/s]= 29.00 \pm 0.17$ and a EM-weighted thermal energy $kT_{\rm av}=0.59 \pm 0.07$~keV. These values are in agreement with our analysis.

The spectral analysis of all available \textit{XMM-Newton} observations, comprising the archival observation of 2014, yields a mean X-ray luminosity $\log L_{\rm X} \ [\rm erg/s] = 29.03 \pm 0.49$, that corresponds to  a X-ray surface flux of $\log F_{\rm X,surf} [\rm erg/cm^2/s] = 6.32 \pm 0.05$.

\subsection{Search for an X-ray cycle}
\label{sec:cycle}
\begin{figure}[!htbp]
\centering
\includegraphics[width=0.5\textwidth]{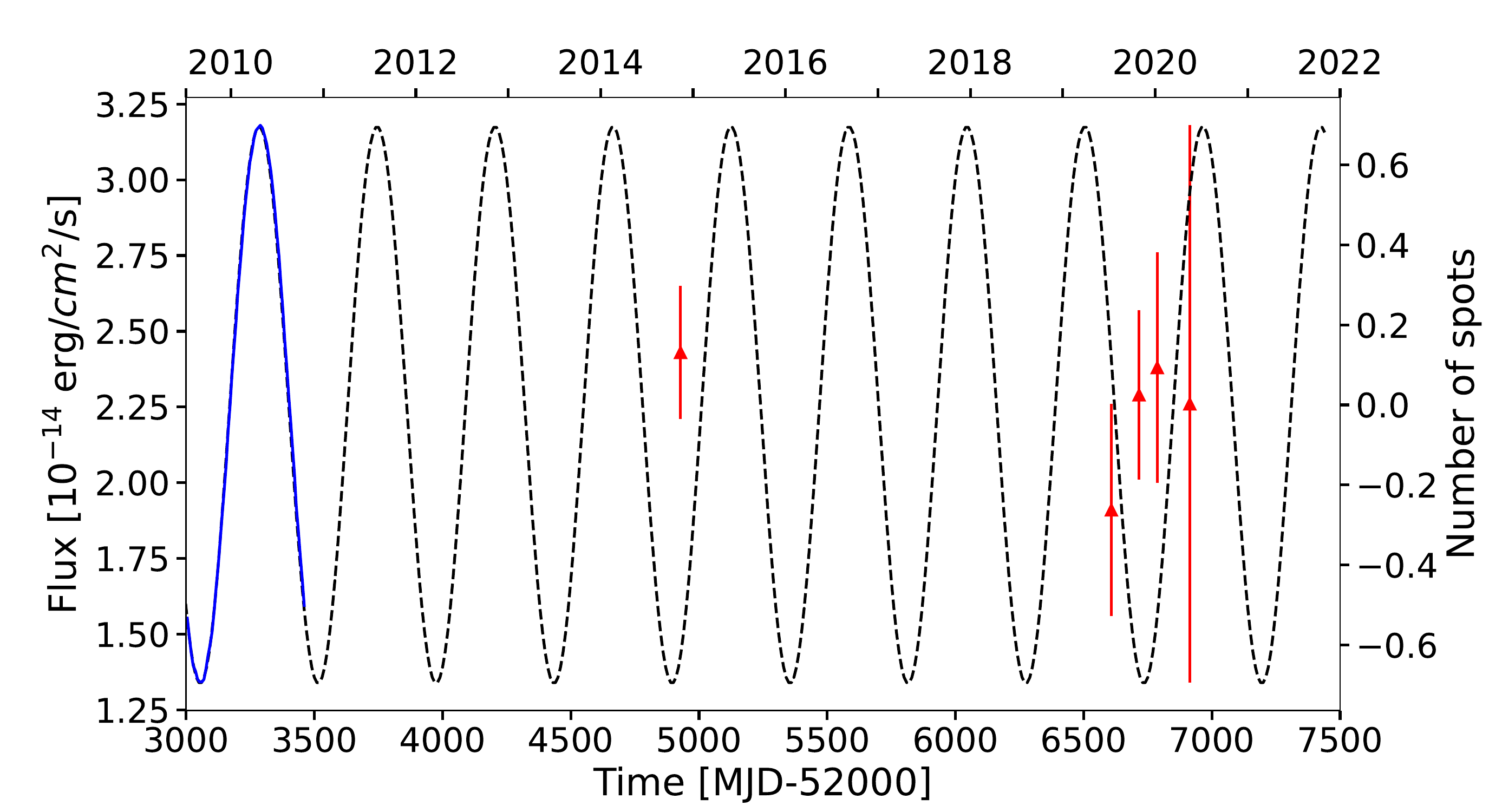}
\caption{Long-term X-ray lightcurve of Kepler-63. The red triangles are the X-ray fluxes, with the associated errors, calculated from the \textit{XMM-Newton} observations. The blue solid line is the sinuisodal function describing the photospheric cycle (1.27 yr) obtained by \cite{2016ApJ...831...57E}. The dotted black line is our extrapolation of the sine curve.}
\label{fig:lc_phcycl}
\end{figure}

The photospheric cycle of Kepler-63 was revealed from observations with the \textit{Kepler} telescope across 4 years (from 2009 to 2013), during which the variation of stellar spots had been monitored. The first observed minimum and maximum happened on August 2009 and March 2010 respectively (see Fig. 10 in \citealt{2016ApJ...831...57E}). We propagated across time the sinusoidal function, representing the $1.27$-yr photospheric cycle of Kepler-63, that \cite{2016ApJ...831...57E} obtained from the Lomb-Scargle analysis of the number of spots seen in the photosphere \footnote{The number of spots seen in the transit curve of Kepler-63 was scaled to the running mean over a range of five data points (i.e. five stellar spots) and to the residuals of a quadratic polynomial fit to the data. This implied that the final number of spots had non-interger and negative values \citep{2016ApJ...831...57E}.}. We found that the maximum and minimum closest to our X-ray campaign were expected for the end of January 2019 and for mid September 2019. In \ref{fig:lc_phcycl} we show the long-term X-ray lightcurve of Kepler-63 where the solid blue line is the folded sinusoidal function calculated by \cite{2016ApJ...831...57E} and the dotted line is our extrapolation of the photospheric cycle. The red symbols are the X-ray fluxes derived from our monitoring of Kepler-63. Clearly, the uncertainties of the X-ray flux in each \textit{XMM-Newton} observation are larger than the changes of the flux between the observations. Therefore, we can only state that any potential X-ray cycle during our campaign in 2019-2020 had a minimum-to-maximum variation smaller than a factor 2. 

\subsection{Activity level and cycle amplitude of Kepler-63}
\begin{figure}[!htbp]
\centering
\includegraphics[width=0.5\textwidth]{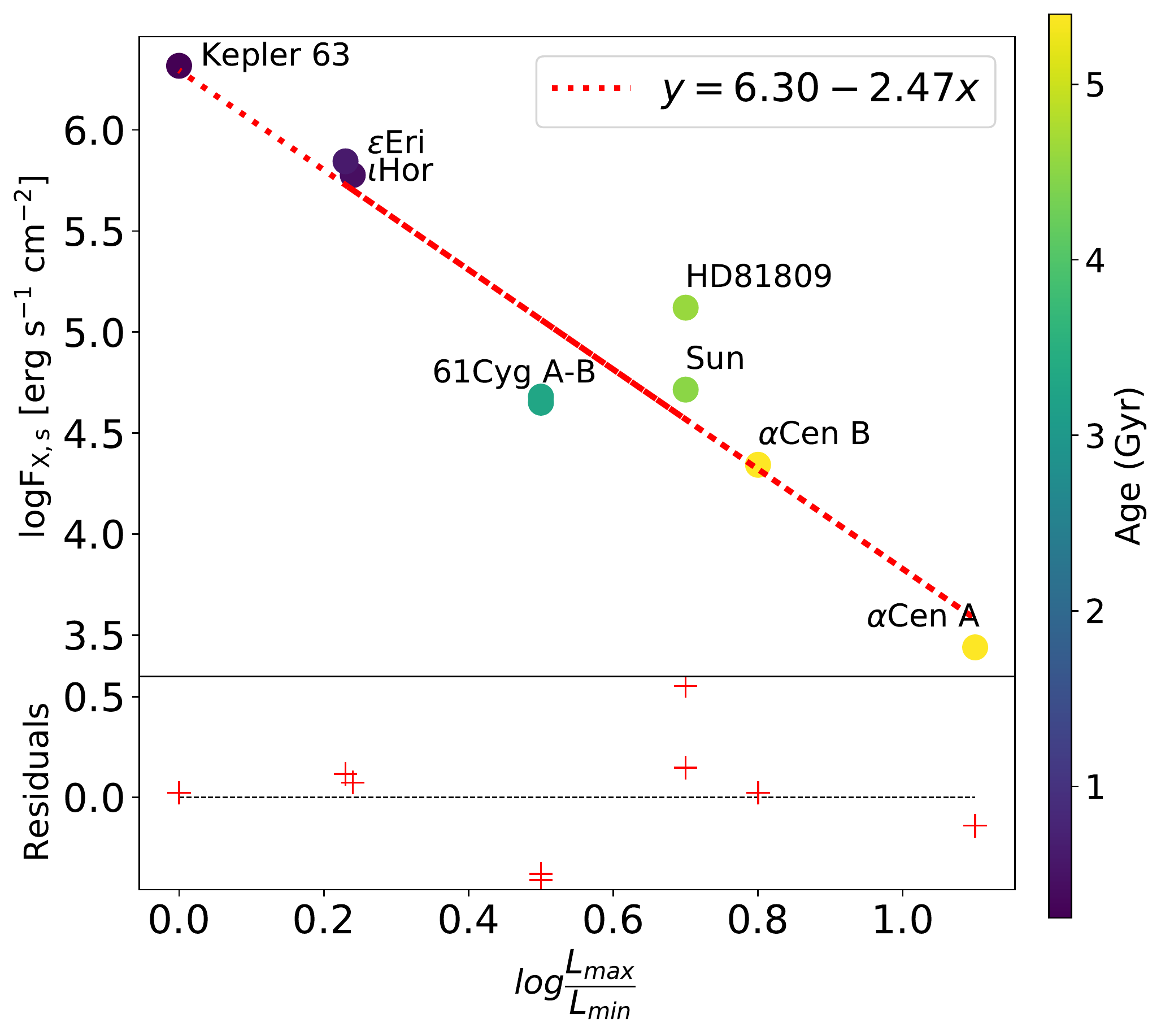}
\caption{X-ray surface flux of all stars with known X-ray cycles as a function of the cycle amplitude i.e. the variation of $L_{\rm X}$ between the maximum and the minimum of the activity cycle. In the plot, also Kepler-63 is included although we detect no X-ray cycle. The color bar denotes the ages of the stars. The dotted red line is the linear regression performed on the data set. The result of the fit is given in the legend of the plot, while the residuals of the procedure are reported in the bottom panel.  }
\label{fig:Lxage}
\end{figure}

In \ref{fig:Lxage} we put Kepler-63 together with stars with known coronal cycle. It can be seen that Kepler-63 is the most active of them, i.e. it presents the highest surface X-ray  flux. 
\ref{fig:Lxage} shows the logarithm of the X-ray surface fluxes ($F_{\rm X,surf}=	L_{\rm X}/4 \pi R_\star$) as a function of the X-ray cycle amplitude ($A=L_{\rm max}/L_{\rm min}$) in log scale, defined as the variation of the X-ray luminosity between the maximum and minimum of the coronal cycle\footnote{The X-ray surface fluxes and the cycle amplitude of each star in the sample are calculated from the X-ray luminosity and the long-term lightcurves published in the literature (the Sun from \citealt{2001ApJ...560..499O}; $\alpha$ Centauri A and B and 61 Cygnus A and B from \citealt{2012A&A...543A..84R}; HD81809 from \citealt{2017A&A...605A..19O}; $\iota$ Horologii from \citealt{2019A&A...631A..45S}; $\epsilon$ Eridani from \citealt{2020A&A...636A..49C}).}. We notice that there is a linear correlation between the log of these two quantities: as the X-ray activity level of the stars decreases (and, in turn, the stellar age increases), the amplitude of their coronal cycles increases.
We interpret this result as due to a decrease in the coverage of the stellar surface with magnetic structures as the star evolves. When the star is young the massive presence of magnetic structures limits the variations of X-ray luminosity during the stellar cycle; when the star is old, the lower coverage of the stellar surface with coronal structures allows the latter to have large changes in their coverage during the cycle, leading to large variations of the X-ray luminosity.
This scenario is consistent with the expectation from dynamo theory where the alternation of the $\alpha$- and $\Omega$-effect causes periodic changes of the number of magnetic structures, and their number is larger for faster rotators. As rotation rate drops with stellar age our color code in \ref{fig:Lxage} can be taken as a code for rotation rate and, in fact, with $P_{\rm rot}=5$~d Kepler-63 is the fastest rotator in the sample.

According to the linear regression on the dataset in \ref{fig:Lxage}, the X-ray surface flux of the stars reaches a maximum value $\log F_{\rm X,surf} \rm{[erg/cm^2/s]} \sim 6.3$ when the stars have no activity cycle (i.e. when the amplitude $\log L_{\rm max}/L_{\rm min} = 0$). This flux, however, can not be considered as an upper limit to the X-ray flux that active stars can reach because X-ray observations have revealed a significant fraction of stars with fluxes up to  $\log F_{\rm X,surf} \rm{[erg/cm^2/s]} \sim 7.5$ \citep{1997A&A...318..215S}. Since we have shown through our modelling of Kepler-63 (summarized in \ref{sec:disc_ms}) that the stellar corona is fully covered with magnetically active structures already at one order of magnitude smaller flux, such high values for $F_{\rm X,surf}$ indicate not only that the coronae of these stars have a 100\% filling factor but also that they must have flares of higher energy than the ones we have inferred to be present on Kepler-63. The presence of high energy flares is a reasonable hypothesis as e.g. pre-main sequence stars have flares with much greater energies (and X-ray fluxes) than solar flares \citep{2021ApJ...916...32G}.

\subsection{Magnetic structures on the corona of Kepler-63 and implications for cycles}
\label{sec:disc_ms}
We quantitatively investigated if the undetected X-ray cycle amplitude is linked to a massive presence of magnetic structures on the corona of Kepler-63, by applying a novel technique based on the study "The Sun as an X-ray star" (\ref{sec:sim_summary}). This technique allows us to indirectly gain information on the magnetic structures that determine the X-ray emission level of Kepler-63. The method builds on the EMDs of the magnetic structures observed by the solar X-ray satellite \textit{Yohkoh} on the corona of the Sun: BKCs, ARs, COs and FLs. The EMDs can be built assuming that different percentages of these magnetic structures determine the total emission. From these EMDs, X-ray spectra of a pseudo-Sun (in the present case, Kepler-63) can then be synthesized as if they were collected with \textit{XMM-Newton}.

In our previous application of this technique to $\epsilon$~Eridani \citep{2020A&A...636A..49C}, the high signal-to-noise ratio of the observations enabled a very detailed description of the coronal temperature structure of the star, outlined by a 3-T model fit to the EPIC/pn spectra. On the other hand for the case of Kepler-63, we can analyse most of its EPIC/pn observations employing a one temperature spectral model,  as seen in \ref{sec:spectra_single}. To increase the signal-to-noise ratio, and thus to use a 2-T spectral model for achieving a more detailed description of the corona of Kepler-63, it was necessary to reduce the X-ray monitoring campaign to only two observational epochs, one describing the corona in 2014 and the other one describing its average behaviour during 2019. This was achieved by simultaneously analysing all observations acquired in that year.

We found that the coronal temperature structure of Kepler-63 is reproduced when we assume time-averaged COs together with a FL EMD. BKCs and ARs are neglected as they have a low surface brightness. Excluding BKCs and ARs does not imply that such regions are not present on the stellar corona. However, the high surface coverage of COs that is necessary to produce an X-ray luminosity as high as that observed leaves only a small percentage of the stellar surface available for BKCs and ARs. The low filling factor of these types of regions together with the fact that they are characterized by a much lower surface brightness in comparison with that of COs and FLs make the contribution of BKCs and ARs to the total X-ray emission of the star negligible with respect to that of COs.

For the flaring contribution to the total EMD, we tested two scenarios: in the first one, we assumed that a distribution of flares from (GOES) Class C to Class X, from their rise to their decay, are present in the corona of Kepler-63, re-conducting its behaviour to a standard flaring Sun. In the second scenario, we assumed that only lower-energetic flares, i.e. from Class C (the least energetic flaring events of our sample) to Class M, are present. This was necessary because, as shown in \ref{fig:ratio}, the first scenario can not reproduce the X-ray temperatures observed in Kepler-63.

For the case of the standard flaring EMD, we examined the effect of scaling the simulated EMDs to two different sizes of the corona: in a first approach we adopted the literature value for the radius of Kepler-63, neglecting the extension of its corona above the chromosphere. In a second test we considered the possibility that the corona of Kepler-63 significantly extends above its surface, and in particular that the extension is $30\%$ of the stellar radius ($R_\star = 0.91 R_\odot$). This hypothesis is motivated by solar observations, from which it is evident that the corona of the Sun is spread above the surface up to $20-30\%$ of its radius \citep{2007ApJ...661..532D}. 

\ref{fig:bestMod1} and \ref{fig:bestMod2} show that, due to the large errors associated to each best-fitting parameter for both observations and synthetic spectra (black and red symbols), both tested scenarios (i.e. the EMDs with different flaring components and different sizes of the corona) can reproduce the spectra of Kepler-63.
The parameter $\sigma$, defined in Eq.\,\ref{eq:crit}, is $\sim 1$ when the selection criterion is run over the first scenario, i.e. the EMDs composed of all classes of FLs plus a $30\%$ coronal extension. In the second tested case, i.e. the EMDs with Class C and M FLs and no extension of the corona, $\sigma$ results to be $\sim 0.6$. Thus, we choose this latter model as the preferred one. For this scenario we find that the magnetic structures cover the whole corona, i.e. the filling factor is $100\%$ for both X-ray epochs of Kepler-63 (\ref{fig:bestMod2}). The complete coverage of the corona of Kepler-63 strengthens the hypothesis that the amplitude of an eventual X-ray cycle is significantly reduced by the massive presence of coronal structures in all phases of the cycle.

 As with this simulation we can reproduce the observations of Kepler-63 without considering the thickness of the corona and, thus, we underestimate the extension of the corona above the stellar surface, the filling factors of the magnetic structures can be taken for upper limits. 
If we consider the corona to have an extent of $30\%$ of the stellar radius covered with the same types of structures (i.e. Class C and M flares and COs), the surface coverage with magnetic structures would be reduced by a factor $1.7$ and only $\sim  60\%$ of the corona would result filled by COs and FLs.
Most likely, the regions not covered with COs and FLs may be filled with BKCs and ARs. This, however, has a negligible impact on the EMD and X-ray emission as explained above. The lower coverage fraction of the extended corona would imply that a change of the X-ray luminosity by a factor of two is possible during the cycle of Kepler-63, as almost half of the corona would be free for new magnetic structures to rise and evolve during a potential cycle maximum. Although our data do not suggest the presence of an X-ray cycle, a factor 2 variation can not be excluded given the large uncertainties of the individual flux measurements (see \ref{fig:lc_phcycl} and \ref{sec:cycle}).

\subsection{Nature of the flares on Kepler-63}
\label{sec:flares}
Another intriguing characteristic we face is related to the nature of the magnetic structures we used for modelling the corona of Kepler-63.
To reproduce best the coronal structure of Kepler-63, the hottest component of the total EMDs needs to be at lower energy than the typical solar flare EMD (which indicatively has temperatures around $7-10$~MK). To represent such energies, we made use of low- and mid-energy flares from Class C to M, but omitting the more energetic Class X events. %We can, therefore, formulate two hypotheses. 
Since young solar-like stars show lower metallicity than the Sun, the radiative cooling is less efficient. This in turn implies that the evolution of flares is slowed down and the events are on average seen at lower temperatures. Thus, it is plausible that overall low- and mid-energetic flares are dominant on Kepler-63, and such a scenario is supported by our models.

We stress that we are exploring a quasi-stationary component of the corona of Kepler-63 resulting from the superposition of several unresolved flares. From the canonical power-law distribution, the frequency of higher-energy flares (such as, for instance, Class X flares) is expected to be lower than that of low-to-mid energetic flares and, therefore, they would not contribute significantly to the quasi-stationary component. High-energy flares may be revealed as short-term variability in the X-ray lightcurves. As a rough estimate, we can assume that a flare would have been detected in the lightcurve if it was $2\delta$ above the mean count rate. Here $\delta$ is the standard deviation, and for Kepler-63 we find $\delta = 0.01 \ \rm cts/s$. Using the rate-to-flux conversion factor obtained from the EPIC/pn spectral analysis ($CF = 1.2 \cdot 10^{-12} \rm erg/cm^2/cts$) such a 2$\delta$ event corresponds to a flux amplitude of $2.5 \cdot 10^{-14} \rm erg/cm^2/s$, and the corresponding peak flare flux at the surface of Kepler-63 is $F_{\rm X,surf,2\delta} = 2.3 \cdot 10^6 \rm  erg/cm^2/s$. The total surface flux of Kepler-63's quiescent X-ray emission (that includes the quasi-stationary flaring component) is $2.1 \cdot 10^6 \ \rm erg/cm^2/s$. This means that a 2$\delta$ event has $1.1\times$ the total surface flux of the star, i.e. only events that provide an X-ray flux that is higher than that of the whole rest of the star's corona can be identified in the EPIC/pn lightcurve. Such events have $45\times$ the surface flux of an X9 Class flare\footnote{This estimate does not consider the different energy bands of \textit{XMM-Newton}/EPIC and the GOES flare classification; see below for an estimate of the conversion.}. No clear flare signatures are seen in \ref{fig:lc}, but the strong variability between two subsequent 1000-bins could, in principle, be caused by such high-energy flares. Judging from the lightcurves, the rate of such events would be one every 2-3 ksec with a duration of $\sim$ 1 ksec (one bin) each. In this interpretation, the mean count rates indicated in \ref{fig:lc} would not be representing the quiescent activity level. We defer a more quantitative analysis of flare detection thresholds to a future work. In the following we estimate the number of Class X flares on Kepler-63 on the basis of the frequency distribution of flare energies.

The coronal filling factor for Class X flares can be estimated from the FFD power-law. In our analysis we have normalized the FFD to the smallest \textit{Yohkoh} flare which is of Class C5.8. With the power slope $\alpha = 1.57$ we infer a flare frequency for Class X9 flares\footnote{We arbitrarily choose the X9 flare for our calculation since this is the largest flare in the \textit{Yohkoh} sample \citep{2001ApJ...557..906R}; a more accurate treatment should take account of the continuous flare distribution.} by $7.2 \cdot 10^{-4}$ smaller than that of Class C5.8. Our modelling of the X-ray spectra of Kepler-63 in \ref{sec:mod2} has yielded a filling factor with FL of $\sim 13.4\%$ of the stellar surface. This number refers to Class C5.8 flares. Accordingly, the filling factor for Class X9 flares is $ ff_{\rm X9} \sim 9.7 \cdot 10^{-5}$. With a reasonable assumption on the coronal area covered by one such flare we can estimate the number of Class X9 flares present on the surface of Kepler-63 at any time. We carry out the calculation assuming that the area covered by a Class X9 flare on the corona is $ A_{\rm f} = 3 \cdot 10^{19} \ cm^2$. This value is directly derived from the \textit{Yohkoh} images for the X9 flare of the sample which is our reference point \citep{2001ApJ...557..906R}. We then scale this result to the total surface area of Kepler-63. The number of Class X9 flares on Kepler-63, $ N_{\rm X9}$, is then given by the ratio of the filling factor $ ff_{\rm X9}$ and the fractional surface area of one such event, $ A_{\rm f}/4\pi R_\star ^2$. We find $ N_{\rm X9} \sim 0.16$. This means bright flares are rare events on Kepler-63, justifying a posteriori the omission of Class X flares in our EMD modelling on timescales consistent with those of observations.

We note that several uncertainties contribute to this calculation of the flare number and energetics. In particular, the number of X9 Class flares depends strongly on the area covered by such an event. The \textit{Yohkoh} X9 flare from \cite{2001ApJ...557..906R} was a so-called complex flare, characterized by an arcade of loops. The value for the flare area derived by \cite{2001ApJ...557..906R} for this event is consistent with the areas inferred by \cite{2008ApJ...674..530A} for GOES Class M and Class X flares. Our above estimate for the flare number, therefore, is likely to be representative for solar flares. 

A second uncertainty in our modelling is the slope of the FFD distribution for which we used the \textit{Yohkoh} value, $\alpha=1.57$. If, instead, we assume a steeper FFD ($\alpha =2.0$, based on observations of stellar flares in the optical, see e.g. summary by \citealt{2021arXiv211209676I}), the conditions are even more relaxed, yielding flare numbers that are lower by approximately a factor 10.

Finally, as mentioned in footnote 7, we are mixing numbers derived from different instruments. The flare classification is based on GOES $1-8 \ \AA$ flux while our analysis uses data from the \textit{Yohkoh}/SXT. \cite{2008A&A...488.1069A} determined the scaling between flare emission measures observed with these two instruments to be $ EM_{\text{Yohkoh}} \sim 0.5 \cdot EM_{\text{GOES}}$. The factor two uncertainties that results if a similar relation holds for the flux values of \textit{Yohkoh} and GOES is clearly smaller than the uncertainties caused by the other assumptions described above.

To summarize, a calculation based on the properties of a Class X9 flare observed with \textit{Yohkoh} shows that the number of such bright flares present in the corona of Kepler-63 is negligible, consistent with the requirements from our "The Sun as an X-ray star" modelling of the X-ray spectrum and with the non-detection of such events in the X-ray lightcurve of the star. However, for other plausible flare parameters a higher number of X9-type events can be obtained, and such a scenario can not be excluded by the X-ray lightcurve that has too poor a signal-to-noise ratio to distinguish individual Class X flares.

\section{Conclusions and outlook}
With our simulations and their comparison to the observed \textit{XMM-Newton} spectra we have identified the types of solar magnetic structures that best reproduce the coronal X-ray emission of Kepler-63. We found that modifications are required as to the presence and relative importance of the different magnetic features with respect to those that dominate the X-ray emission of the Sun. In our earlier analogous study of another young solar-like star, $\epsilon$~Eridani \citep{2020A&A...636A..49C}, we also had to adapt the solar magnetic structures to be able to fit the observations of the stellar corona, but in different ways from what we require for Kepler-63. We conclude that stellar coronae come in a variety of physical characteristics.

The non-detection of an X-ray cycle on Kepler-63 is compatible with
the complete coverage of its corona with magnetic structures as we have derived from
our modelling. The correlation between cycle amplitude and X-ray surface flux that we find 
for the sample of stars with X-ray cycles suggests a limiting X-ray brightness 
for stars that can sustain cycles of $\log{F_{\rm X, surf}}\,{\rm [erg/cm^2/s]} \sim 6.3$,
the value of Kepler-63. Stars with much higher values of X-ray surface flux
than Kepler-63 are known (see e.g. \citealt{1997A&A...318..215S}), and we speculate that these stars must 
present higher energy flares in order to reach their observed X-ray radiation output.

The modelling is likely also affected by the quality of the observations, e.g. the signal-to-noise ratio of the stellar X-ray spectra and the range of flares included in the \textit{Yohkoh} EMDs. Future work should address these points by investigating additional stellar X-ray observations with the technique of "The Sun as an X-ray star" and by closing holes in the solar flare statistics.

\begin{acknowledgements} 
We thank the anonymous referee for the useful input that improved our final discussion. 
MC acknowledges financial support by the Bundesministerium f\"ur Wirtschaft und Energie through the Deutsches Zentrum f\"ur Luft- und Raumfahrt e.V. (DLR) under the grant FKZ 50 OR 2008. 
S.O. acknowledges financial contribution from the agreement ASI-INAF n.2018-16-HH.0.
This work is based on observations obtained with\textit{ XMM-Newton}, an ESA science mission with instruments and contributions directly funded by ESA Member States and NASA.
\end{acknowledgements}

\bibliographystyle{aa} 
\bibliography{kep_20}

\begin{appendix}
\label{appendix:shot_lc}
\section{EPIC/pn lightcurve of the background detector}
Here, the lightcurve of the background detector of the EPIC/pn for each observation of Kepler-63 are shown. In each plot, the red lines and red intervals denote the GTIs chosen in each lightcurve as explained in \ref{sec:obs_log}.
\begin{figure}[!htbp]
\begin{minipage}{\textwidth}
\subfloat{\includegraphics[width=0.5\textwidth]{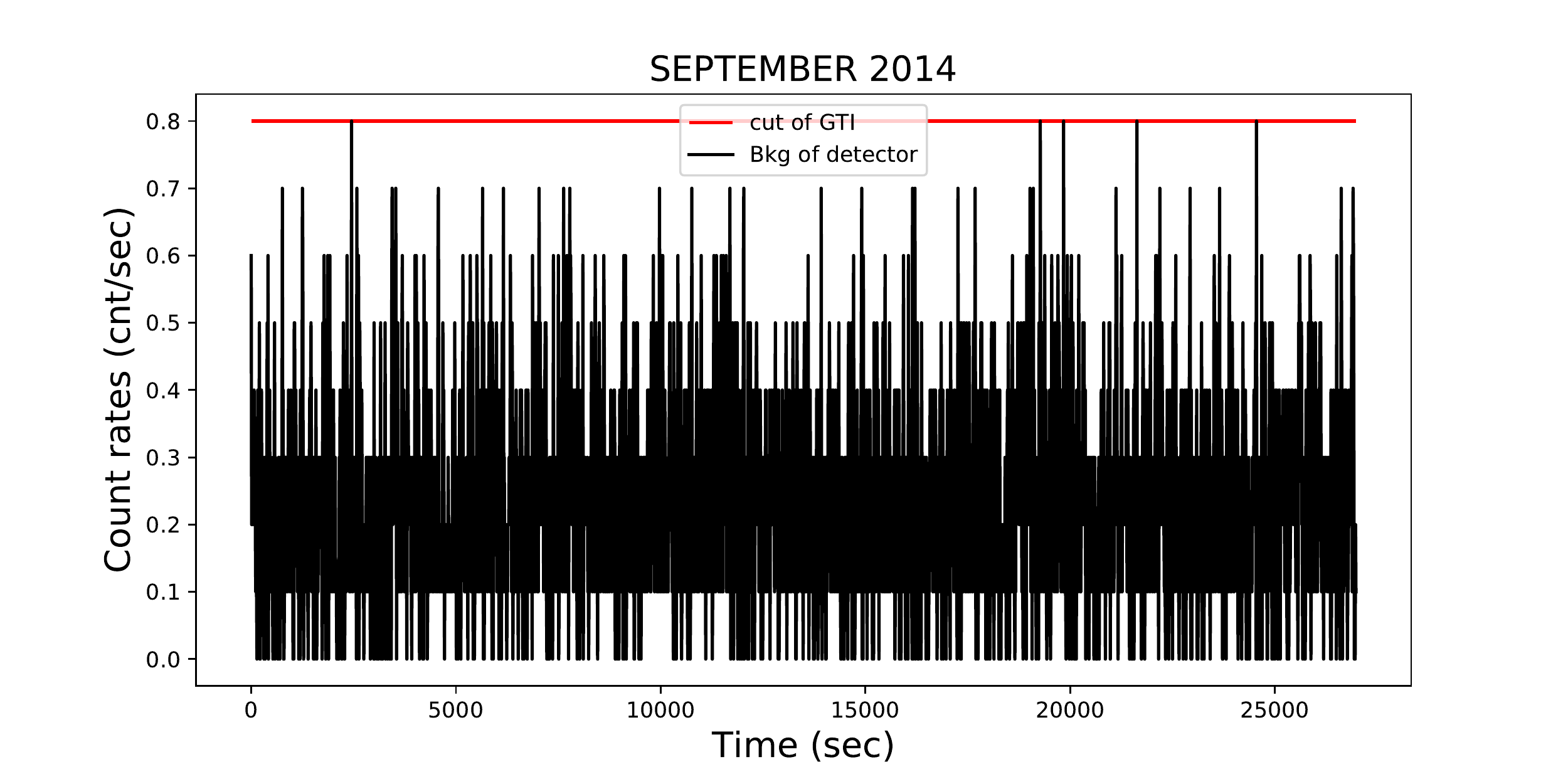}}
\subfloat{\includegraphics[width=0.5\textwidth]{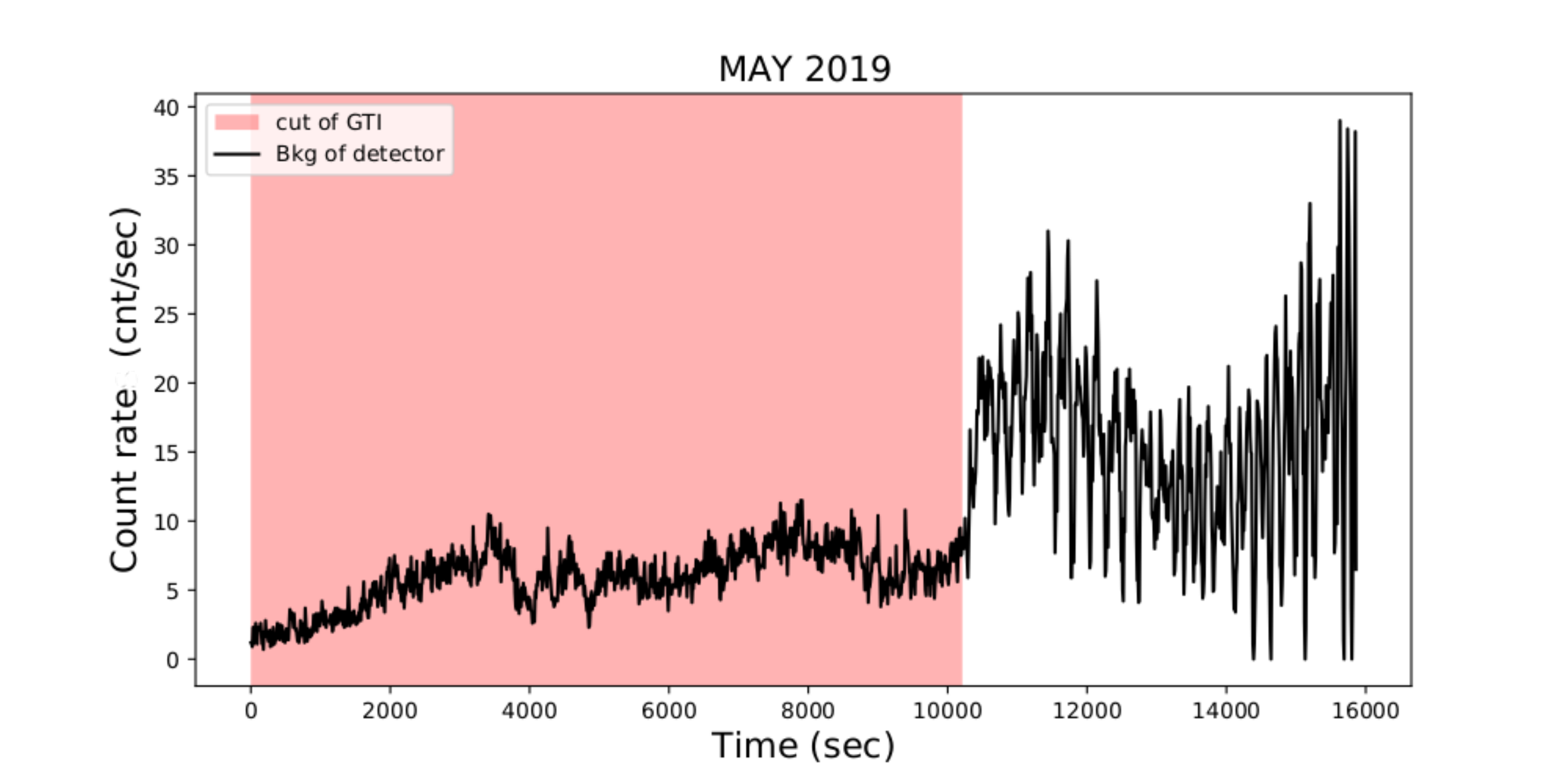}}\\
\subfloat{\includegraphics[width=0.5\textwidth]{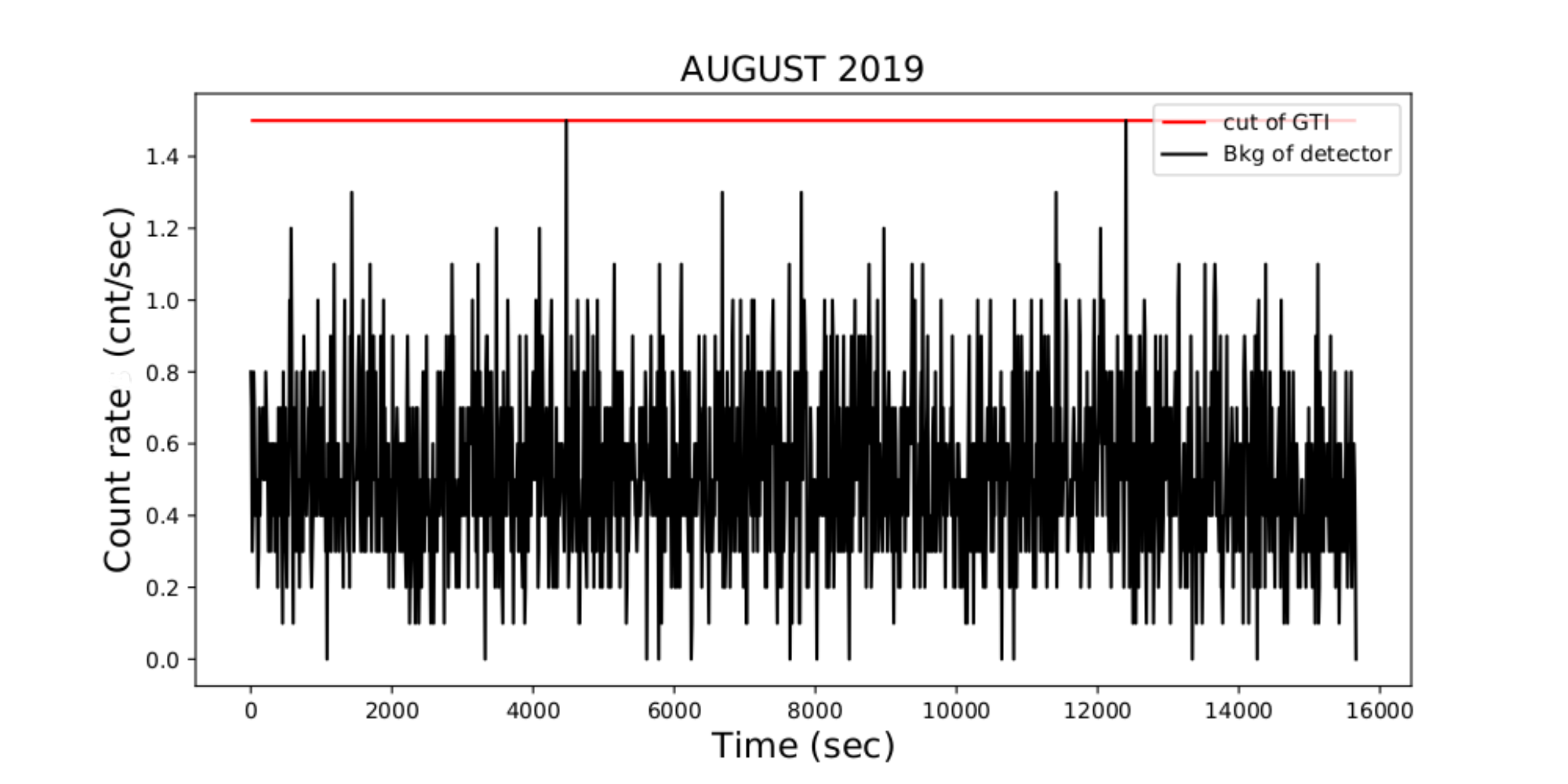}}
\subfloat{\includegraphics[width=0.5\textwidth]{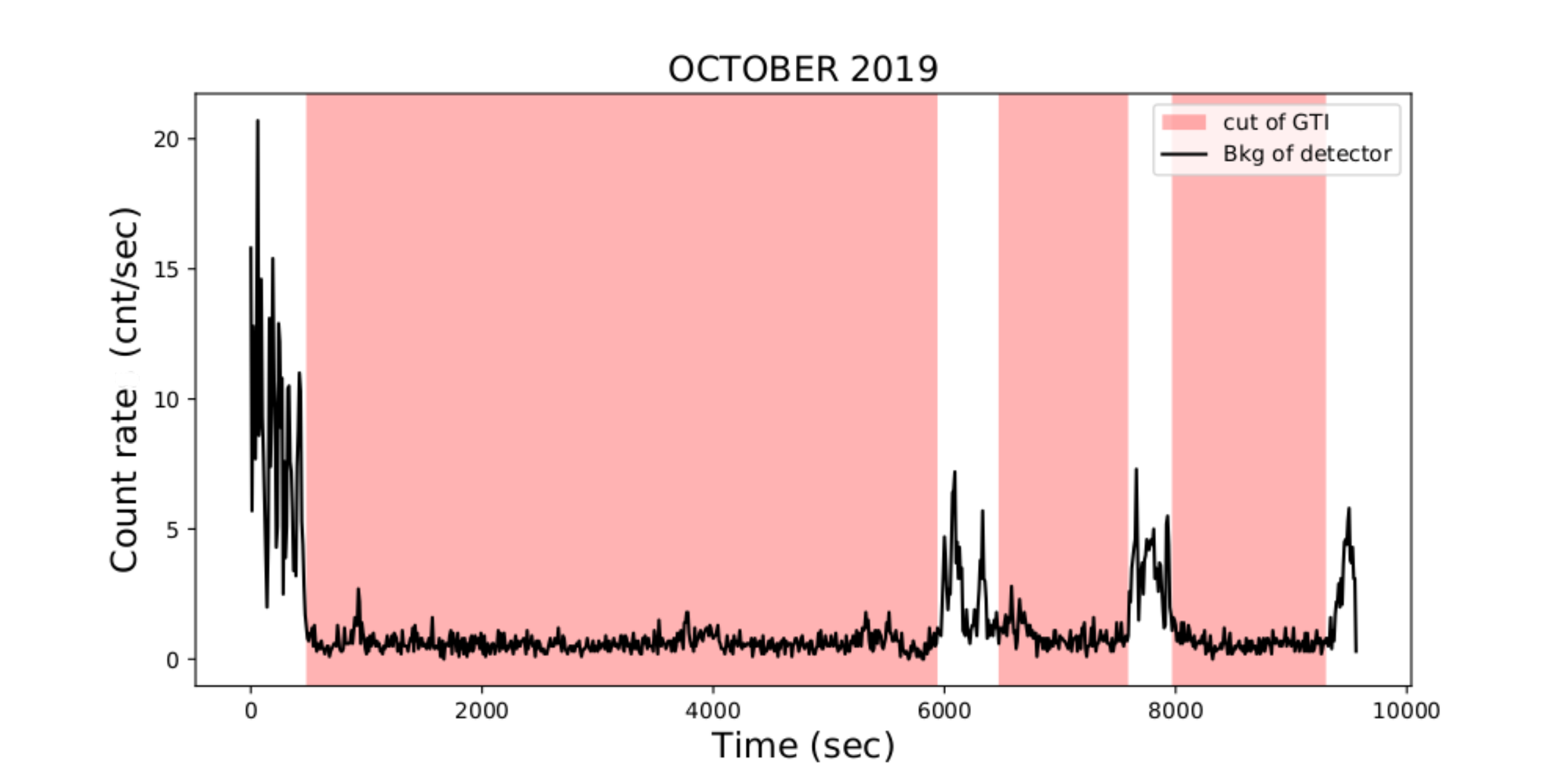}}\\
\subfloat{\includegraphics[width=0.5\textwidth]{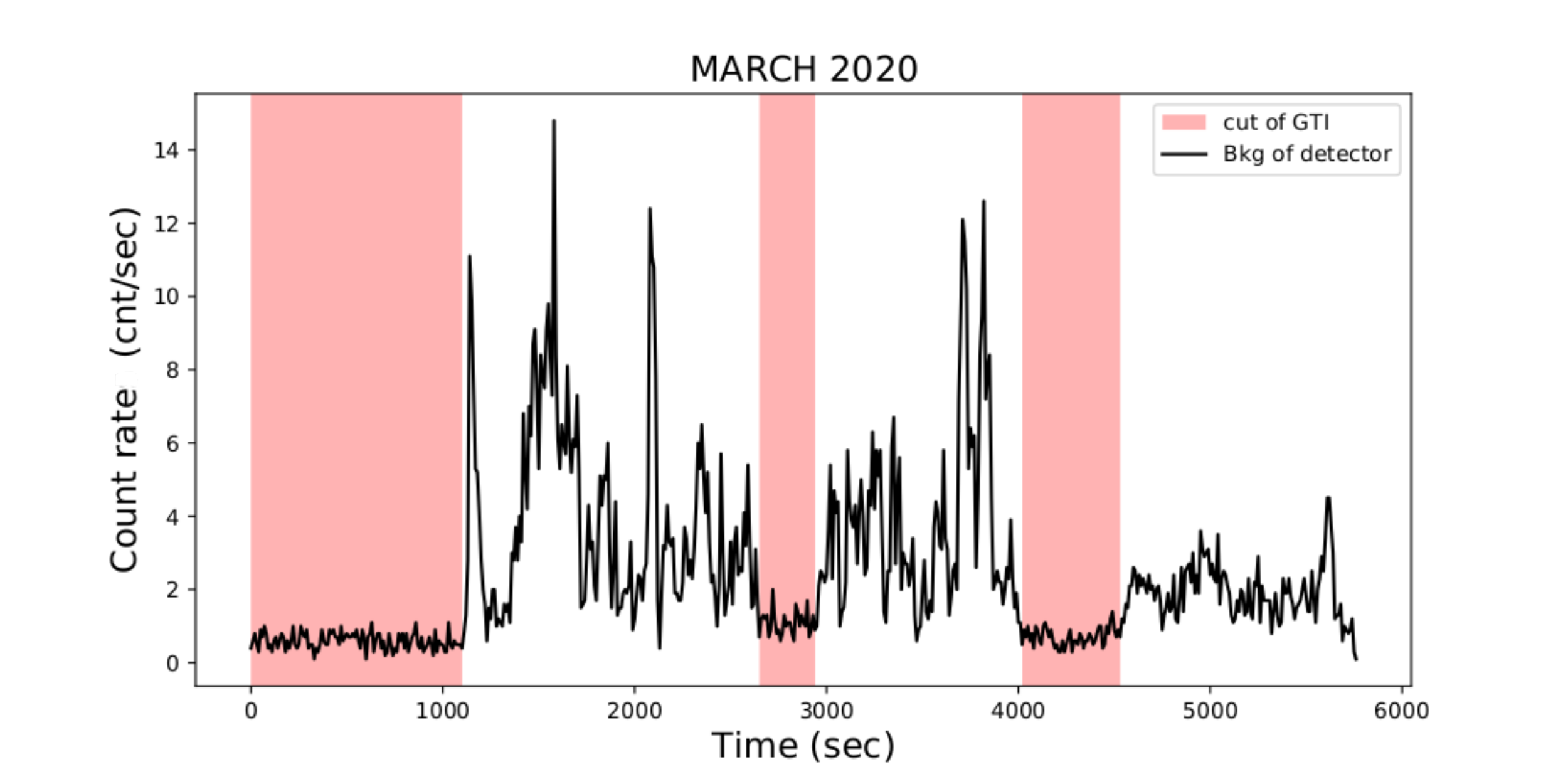}}\\
\caption{Lightcurves of the full-detector background of EPIC/pn. The red lines and the red intervals show where the GTIs were chosen.}
\label{fig:bkg_lc}
\end{minipage}
\end{figure}
\clearpage

\section{Spectral analysis}
The EPIC/pn spectra of Kepler-63 are shown in this section. The solid red line is the spectral model (APEC) used for fitting the observations. The bottom panels in each plot are the residuals of the fitting. 
For the observations of 2014, of August 2019 and the simultaneous fit of 2019 observations, a two temperature spectral model is adopted. In all other cases, only one thermal component is used, with the exception of the observation of March 2020. As explained in \ref{sec:spectra_single}, this latter observation suffers from low photon statistics, and it is not possible to carry out a meaningful fit.
Values of the best-fitting parameters are found in \ref{tab:1T}. 
\begin{figure}[!htbp]
\begin{minipage}[b][19.8cm]{\textwidth}
\centering
\subfloat{\includegraphics[width=0.5\textwidth]{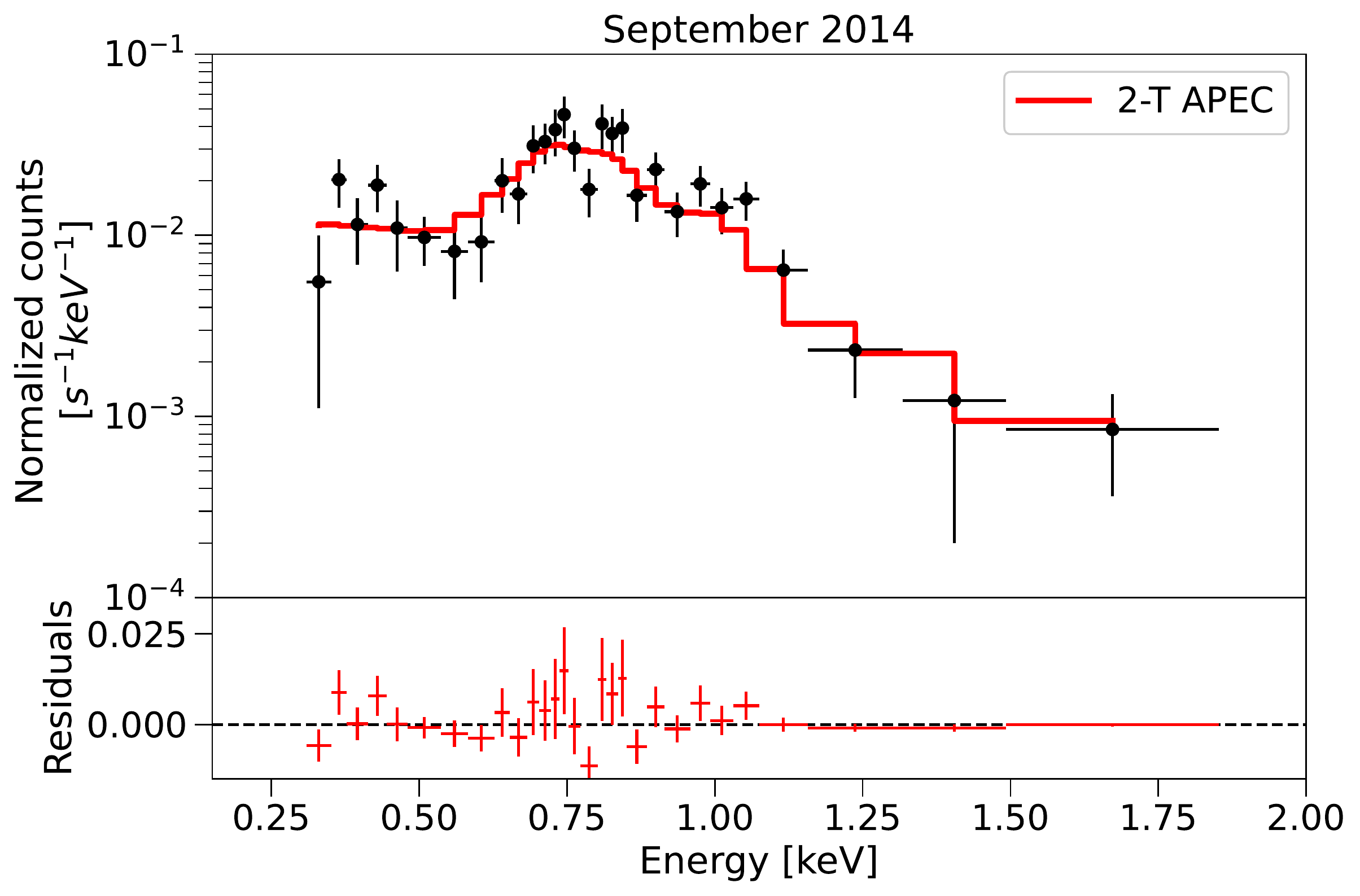}}
\subfloat{\includegraphics[width=0.5\textwidth]{ 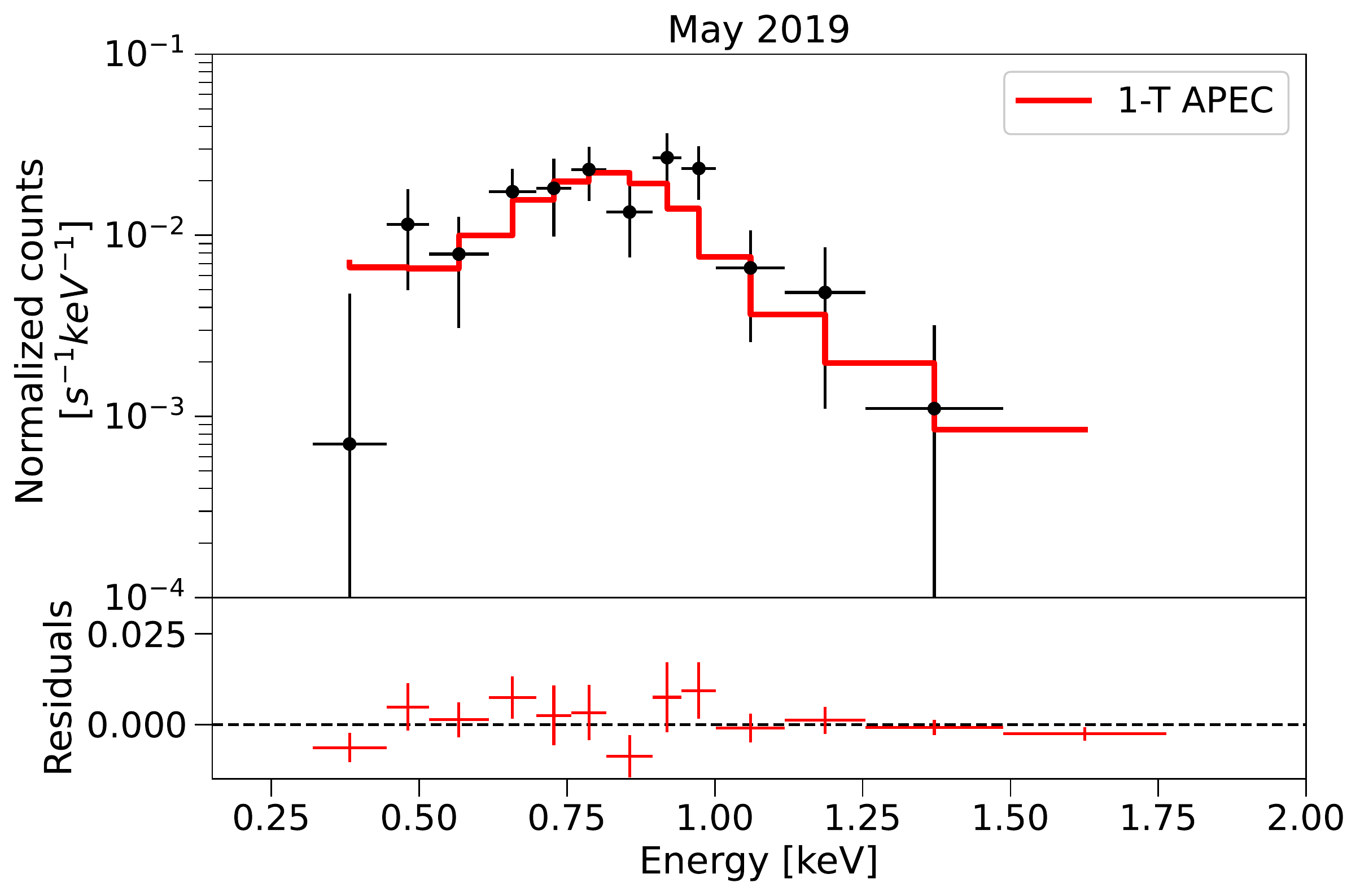}}\\
\subfloat{\includegraphics[width=0.5\textwidth]{ 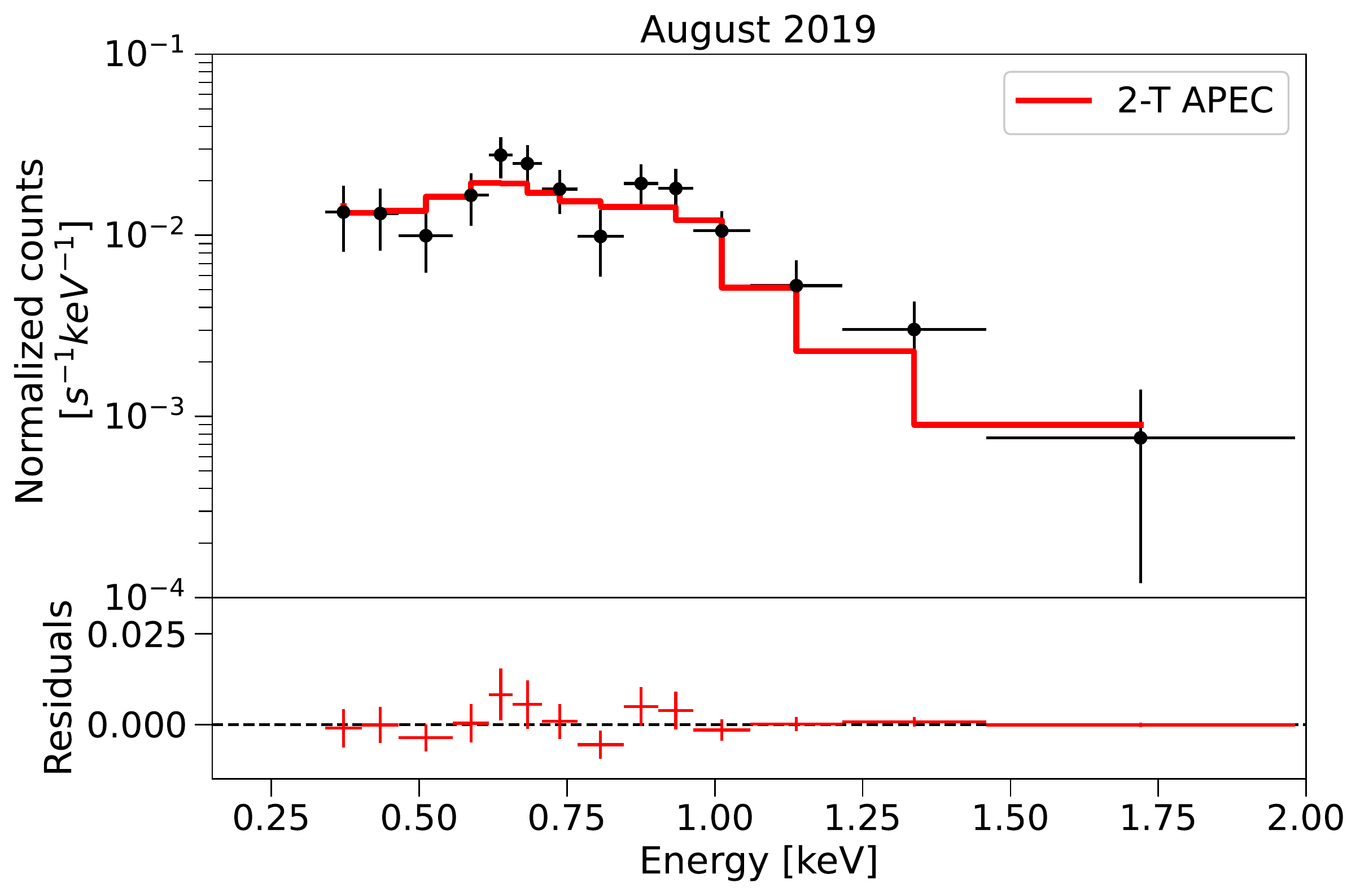}}
\subfloat{\includegraphics[width=0.5\textwidth]{ 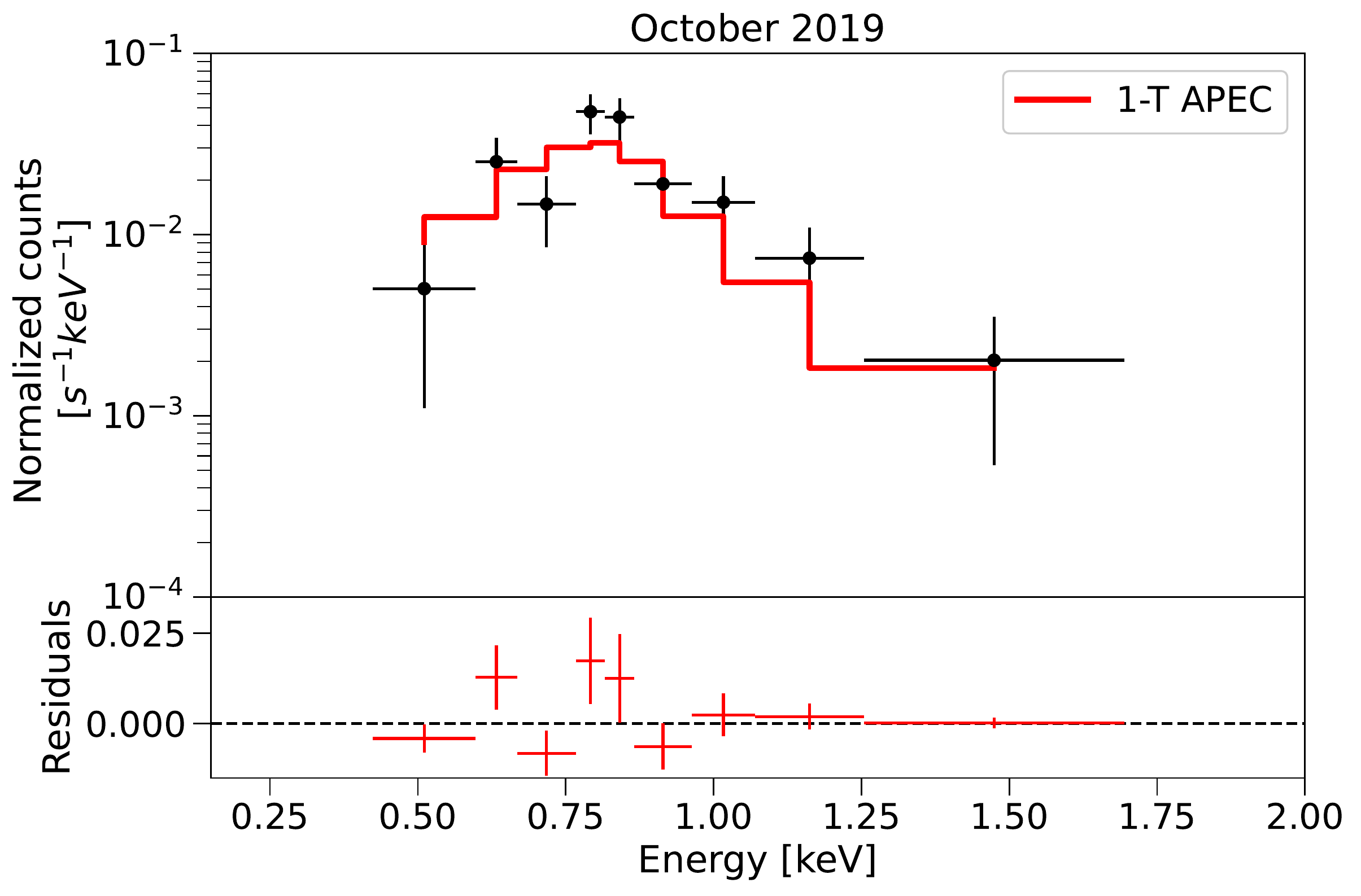}}\\
\subfloat{\includegraphics[width=0.5\textwidth]{ 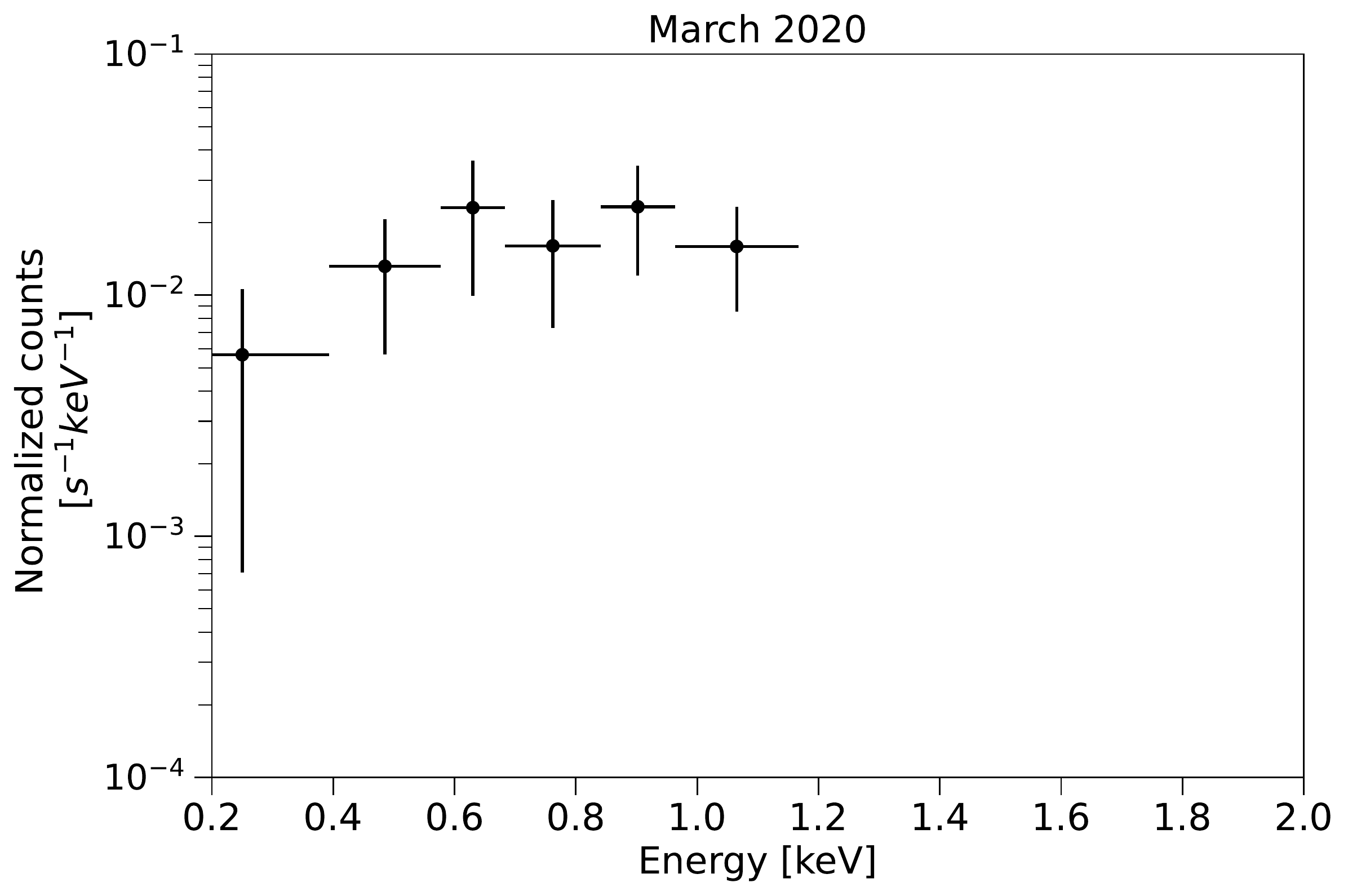}}
\subfloat{\includegraphics[width=0.5\textwidth]{ 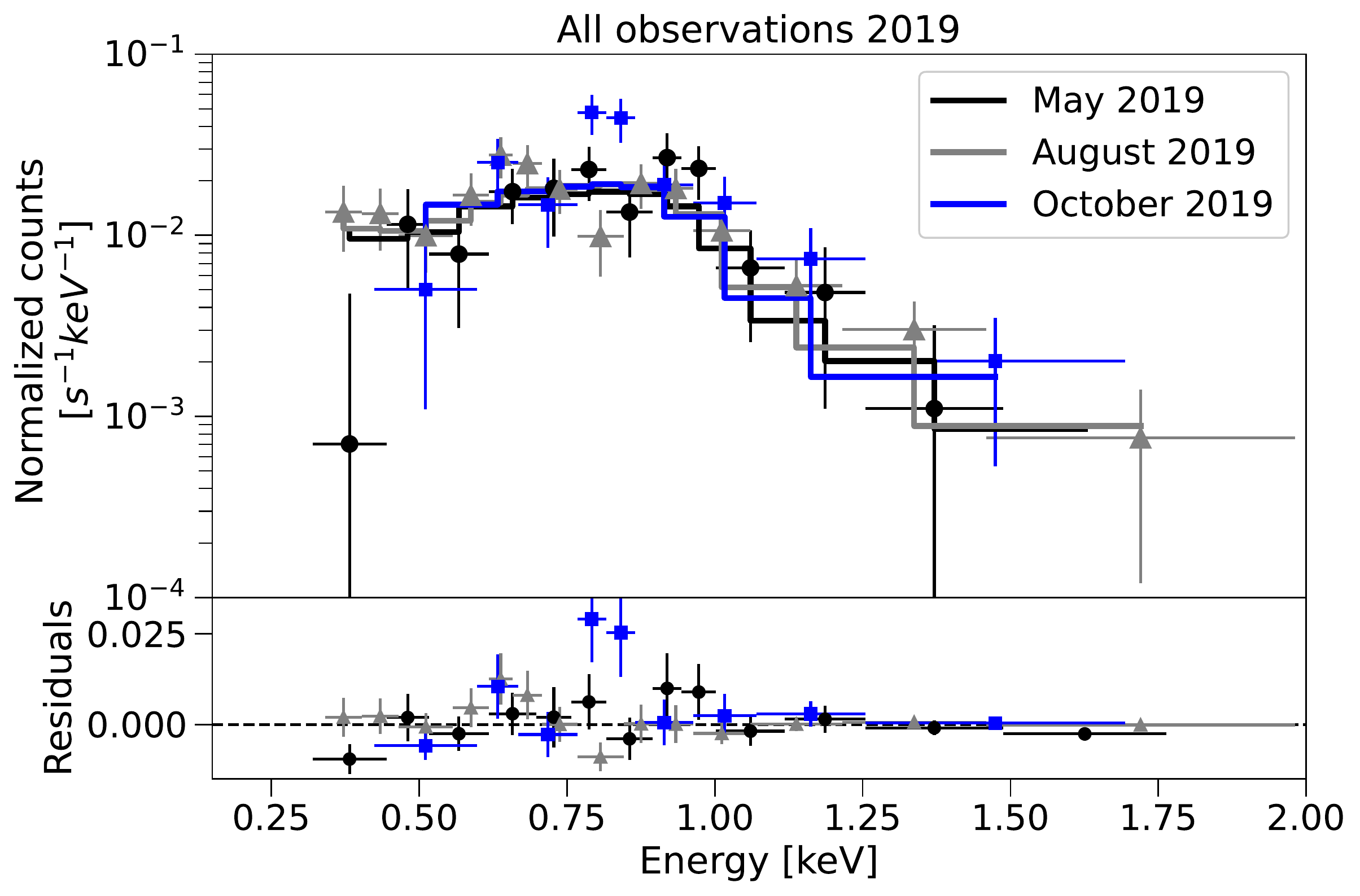}}\\
\caption{EPIC/pn spectra of Kepler-63. The solid line is the best-fitting spectral model. The bottom panels are the reduced $\chi^2$ residuals of the fitting. }
\label{fig:fit_spec}
\end{minipage}
\end{figure}

\end{appendix}

\end{document}